\documentclass[11pt]{article}
\pdfoutput=1
\usepackage[margin=1in]{geometry} 
\usepackage{amsmath,amsthm,amssymb,amsfonts,titling,bbm, titlesec, bm, physics, enumitem, accents, xcolor, setspace, fancyhdr, natbib, geometry,
	pdflscape}
\usepackage{graphicx}
\usepackage{float}
\usepackage[font=footnotesize,labelfont=bf]{caption}
\usepackage{array}
\usepackage[title]{appendix}
\usepackage{afterpage}
\usepackage{listings}
\usepackage{prodint}
\usepackage{algorithm}
\usepackage{algpseudocode}
\usepackage{algorithmicx}
\usepackage[hidelinks]{hyperref}
\usepackage{multirow}
\usepackage{booktabs}

\newlength{\continueindent}
\usepackage{etoolbox}
\makeatletter
\newcommand*{\ALG@customparshape}{\parshape 2 \leftmargin \linewidth \dimexpr\ALG@tlm+\continueindent\relax \dimexpr\linewidth+\leftmargin-\ALG@tlm-\continueindent\relax}
\apptocmd{\ALG@beginblock}{\ALG@customparshape}{}{\errmessage{failed to patch}}
\makeatother

\algnewcommand\algorithmicstack{\textit{Stack:}}
\algnewcommand\Stack{\item[\algorithmicstack]}
\algnewcommand\algorithmicchoosetimegrid{\textit{Choose time grid:}}
\algnewcommand\Choosetimegrid{\item[\algorithmicchoosetimegrid]}
\algnewcommand\algorithmicchoosetimebasis{\textit{Choose time basis:}}
\algnewcommand\Choosetimebasis{\item[\algorithmicchoosetimebasis]}
\algnewcommand\algorithmicbuilddiscretedata{\textit{Build data at time $t_i$:}}
\algnewcommand\Builddiscretedata{\item[\algorithmicbuilddiscretedata]}
\algnewcommand\algorithmicfit{\textit{Fit:}}
\algnewcommand\Fit{\item[\algorithmicfit]}
\algnewcommand\algorithmicpredict{\textit{Predict:}}
\algnewcommand\Predict{\item[\algorithmicpredict]}

\newcommand\Algphase[1]{%
	\vspace*{-.7\baselineskip}\Statex\hspace*{\dimexpr-\algorithmicindent-2pt\relax}\rule{\textwidth}{0.4pt}%
	\Statex\hspace*{-\algorithmicindent}\textbf{#1}%
	\vspace*{-.7\baselineskip}\Statex\hspace*{\dimexpr-\algorithmicindent-2pt\relax}\rule{\textwidth}{0.4pt}%
}

\newcommand{\R}{\mathbb{R}}
\newcommand{\I}{\mathbbm{1}}
\newcommand{\matr}[1]{\mathbf{#1}}
\newcommand{\midd}{\,|\,}

\titleformat{\section}{\scshape\bfseries}{\thesection.}{1em}{}
\titleformat{\subsection}{\normalfont\bfseries}{\thesubsection.}{1em}{}

\begin{document}
	\begin{center}
		\textbf{A framework for leveraging machine learning tools to estimate}\\
		\textbf{personalized survival curves}\\ \vspace{0.3cm}
		Charles J. Wolock\textsuperscript{1}, Peter B. Gilbert\textsuperscript{3, 2}, Noah Simon\textsuperscript{2} \& Marco Carone\textsuperscript{2,3}
		\\ \vspace{0.3cm}
		\footnotesize{\textsuperscript{1}Department of Biostatistics, Epidemiology and Informatics, University of Pennsylvania\\
			\textsuperscript{2}Department of Biostatistics, University of Washington\\
			\textsuperscript{3}Vaccine and Infectious Disease Division, Fred Hutchinson  Cancer Center}
	\end{center}
	
	\noindent \textbf{Abstract:} The conditional survival function of a time-to-event outcome subject to censoring and truncation is a common target of estimation in survival analysis. This parameter may be of scientific interest and also often appears as a nuisance in nonparametric and semiparametric problems. In addition to classical parametric and semiparametric methods (e.g., based on the Cox proportional hazards model), flexible machine learning approaches have been developed to estimate the conditional survival function. However, many of these methods are either implicitly or explicitly targeted toward risk stratification rather than overall survival function estimation. Others apply only to discrete-time settings or require inverse probability of censoring weights, which can be as difficult to estimate as the outcome survival function itself. Here, we employ a decomposition of the conditional survival function in terms of observable regression models in which censoring and truncation play no role. This allows application of an array of flexible regression and classification methods rather than only approaches that explicitly handle the complexities inherent to survival data. We outline estimation procedures based on this decomposition, empirically assess their performance, and demonstrate their use on data from an HIV vaccine trial.\vspace{0.4cm}	
	\\ ~\ 
	\noindent \textit{Keywords:} Censoring, nonparametric, survival analysis, truncation

	\section{Introduction}\label{sec:introduction}
	In the analysis of time-to-event data, the conditional survival function is a key quantity of interest. Within the biomedical field, the conditional survival function, which describes the distribution of an outcome variable conditional on a set of covariates, is especially relevant for prediction. For example, the survival function of a clinical outcome, such as death or disease recurrence, conditional on baseline characteristics may allow a clinician to better understand a patient's medical prognosis. The conditional survival function also appears as a function-valued nuisance parameter in nonparametric and semiparametric survival analysis problems (see, for example, \citealp{Diaz2019} and \citealp{Westling2023}). 
	
	Typically, the analysis of survival data is complicated by the fact that the data are subject to censoring, truncation, or both, depending on the study design. In prospective studies, participants are sampled from the population of interest and followed over time, ideally until experiencing the event of interest. However, loss to follow-up or study termination may preclude observation of the event time. Participants who do not experience the event of interest during follow-up are considered right-censored. Additionally, individuals who have already experienced the event at study initiation are not eligible for recruitment. This sampling constraint is referred to as left truncation. Conversely, in retrospective studies, individuals must have already experienced the event in order to be recruited into the study, leading to right truncation. In this design, censoring is generally not a concern. All forms of truncation lead to systematic selection bias.   
	
	There is a substantial literature focused on estimating the conditional survival function, which we review in Section \ref{sec:lit review}. Many existing methods directly or indirectly aim to minimize an empirical risk based on one of two loss functions: the inverse probability of censoring weighted (IPCW) loss for survival function estimation at a single time-point, or the hazard loss for estimation of a discrete-time hazard function \citep{Polley2011}. The IPCW loss requires estimation of the conditional survival function of the censoring variable and does not correct for truncation-induced sampling bias, while the discrete-time hazard loss does not apply to events occurring in continuous time. Methods employing objective functions for risk stratification, such as the Cox partial likelihood \citep{Cox1972}, do not explicitly target survival function estimation but may produce estimates as a byproduct. 
	
	In this article, we consider decompositions of the conditional survival function that allow use of standard loss-based estimation of functionals of the observed data distribution. These decompositions underlie our proposed method, called \emph{global survival stacking}, which involves estimating a small number of binary regression functions using tools neither specially designed to handle censoring nor truncation. The strengths of this approach include:
	
	\begin{enumerate}
		\item it is a general framework in which practitioners can employ any off-the-shelf learner designed for binary regression or classification;  
		\item it can be applied in prospective (left truncation and right censoring) and retrospective (right truncation) settings without assuming a discrete-time process;
		\item it simultaneously yields estimates of both the event time and censoring time conditional survival functions using the same fitted regressions.
	\end{enumerate}
	The article is organized as follows: In the remainder of this section, we review existing methods for conditional survival function estimation. In Section \ref{sec:data structures and identification}, we describe the data structures emerging from survival studies, provide identification results that form the basis for our estimation framework, and propose the global survival stacking procedure. In Section \ref{sec:simulations}, we evaluate the performance of our class of estimators and demonstrate their use on data from the STEP HIV vaccine trial. In Section \ref{sec:discussion}, we provide concluding remarks. Technical details and additional results can be found in the Appendix. Code to reproduce all results is available online at \url{https://github.com/cwolock/stack_supplementary}. We have implemented global survival stacking in the \texttt{R}
	package \texttt{survML}, available on CRAN. 
	
	\subsection{Review of related work}\label{sec:lit review}
	
	Even under right censoring alone, standard regression techniques cannot be directly applied to estimation of the conditional survival function. However, a number of survival-specific approaches have been proposed. Parametric methods, such as exponential or Weibull regression, are straightforward to use and automatically yield inference, but their validity relies on strong distributional assumptions. The most common regression model used to study survival outcomes is the semiparametric Cox proportional hazards model \citep{Cox1972}. Hazard ratio estimates from the estimated Cox model can be combined with an estimate of the baseline cumulative hazard function (e.g., the \citealp{Breslow1972} estimator) to yield a conditional survival function estimate. A common alternative to the Cox model is the accelerated failure time model \citep{Wei1992}, which is usually implemented in a fully parametric manner. Semiparametric implementations \citep{Lin2013} exist but are seldom used because they are complicated and can be unstable \citep{Zeng2007}. If the covariates of interest are low-dimensional and discrete-valued, a stratified Kaplan-Meier estimator \citep{Kaplan1958} may be reasonable. This method breaks down in moderate dimensions, or when the covariates include continuous variables. \citet{Beran1981} introduced a conditional Kaplan-Meier estimator using kernel smoothing. However, kernel-based methods tend to perform poorly as the number of covariates grows. 
	
	Fortunately, machine learning algorithms offer strategies for estimating complex functions of large numbers of covariates in a flexible manner, leading to a proliferation of such methods in survival analysis. This motivates a discussion of the precise objectives of these methods. Estimation of the conditional survival function and risk stratification are distinct tasks that are often conflated. The Cox proportional hazards model and related machine learning methods are based on the partial likelihood. Maximization of the partial likelihood is equivalent to maximization of the expected concordance between risk scores and survival times \citep{Tarkhan2022}. Indeed, the partial likelihood has no dependence on actual event times and relies only on the relative ordering of events. Due to this fact, methods based on the Cox model might be best understood as risk stratification techniques. Likewise, in random survival forests \citep{Ishwaran2008}, trees are built and evaluated based on stratification objective functions that can be evaluated with censored data (e.g., the log-rank test statistic or Harrell's concordance index \citep{Harrell1982}). It is often the case that conditional survival function estimates can be derived from a risk stratification algorithm (e.g., combining the Breslow baseline hazard estimate with a fitted Cox model), but ultimately these are a byproduct rather than the core goal of stratification approaches. 
	
	Alternatively, one may use IPCW to connect a full-data loss function, such as squared-error loss, to a loss that can be evaluated using the observed data \citep{VanderLaan2003}. Defining an observed-data loss function allows the use of many established learning methods, such as random forests and boosting \citep{Hothorn2006}. These methods rely on estimation of the conditional survival function of the censoring variable, which in general is no easier than estimation of the conditional survival function of the outcome. Recently, \citet{Westling2023} framed the conditional event time and censoring survival functions as minimizers of oracle risks, which allows for iterative empirical risk minimization in order to combine multiple candidate estimators in a Super Learner approach \citep{VanDerLaan2007}. This method, termed survival Super Learner, is appealing because unlike usual IPCW loss functions, which are evaluated at a single time-point, it targets the entire survival function and simultaneously provides estimates of the outcome and censoring distributions. However, the candidate learners comprising the Super Learner are limited to existing survival-specific methods, and the risk functions do not account for truncation. 
	
	Another strategy for estimating the conditional survival function relies on the assumption that events occur in discrete time. For discrete time-to-event variables, the hazard function at a single time is a conditional probability whose estimation can be framed as a binary regression problem in terms of the observed data distribution: among those who have not experienced the event by time $t$, what proportion experience the outcome at that time? Reframing survival function estimation in terms of discrete-time hazard functions allows use of a wider array of machine learning algorithms. Estimation of the survival function at time $t$ involves computing the product of one minus the hazard at each time-point up to and including $t$. For some discretization methods, the conditional hazard is estimated at each time based on a separate binary regression \citep{Yu2011}. Another strategy, which has recently been referred to as ``survival stacking" \citep{Craig2021} and which we hereafter call \emph{local survival stacking}, involves estimating a single regression including time as a covariate. This framework dates back at least to work by \citet{Polley2011}. With time discretized, a survival dataset can be transformed into a longitudinal data set, where each individual appears in the data set at each observed event time until exiting the risk set. It is unclear how performance depends on the choice of time discretization. Left truncation is handled by including individuals in the longitudinal data for as many time-points as they remain in the risk set.
	
	Outside of the discrete-time framework, few proposed methods have explicitly viewed conditional survival function estimation in terms of the observed data distribution. One such method uses generative adversarial networks to learn the joint distribution of the observed data \citep{Zhou2022}; however, it is tied to a specific machine learning architecture and does not handle truncation.
	
	\section{Materials and methods}\label{sec:data structures and identification}
	
	\subsection{Ideal data and parameter of interest}\label{subsec:ideal data}
	
	Suppose that $X$ is a vector of baseline covariates taking values in $\mathcal{X} \subset \R^p$, and $T \in (0, \infty)$ is the event time of interest. The ideal data unit is $O^*:= (X, T)$. We use $P^*$ to denote the distribution of $O^*$. In reality, $O^*$ is observed subject to both censoring and truncation, which are determined by the study design. The observed data consist of $n$ independent and identically distributed observations $O_1, O_2,\dots, O_n$ drawn from $P$, the observed data distribution implied by $P^*$. The relationship between $P^*$ and $P$ is determined by the censoring and sampling mechanisms. Our goal is estimation of the conditional survival function of $T$ given $X$, defined as $S(t \midd x) := P^*(T > t \midd X = x)$. Because $T$ is not directly observed, this parameter is not a functional of the observed data distribution. However, with an additional assumption, the conditional hazard function (and through it, the conditional survival function) can be identified. Our method relies on a reformulation of standard identification results in order to write the hazard function in terms of observable regression functions.
	
	Let $\Lambda(t \midd x) := \int_0^t \{1 - F(u^- \midd x)\}^{-1}F(du \midd x)$ denote the conditional cumulative hazard of $T$ given $X=x$ at time $t$, 
	where $F := 1 - S$ is the conditional distribution function of $T$ and $F(u^-) := \lim_{v \uparrow u}F(v)$. Identification of $S$ in full generality requires the use of product integrals \citep{Gill1990} via the mapping
	$S(t \midd x) = \prodi_{u \in (0, t]}\left\{1 - \Lambda(du \midd x)\right\},$ where, for a partition $0 = t_0 < t_1 < \cdots < t_k  = t$ of $(0, t]$ and measure $M$, $
	\prodi_{u \in (0, t]}\left\{1 + M(du)\right\} := \lim_{\max\abs{t_i - t_{i-1}}\to 0}\prod_i \left\{1 + M\left((t_{i-1}, t_i]\right)\right\}$.
	When the mapping $t \mapsto F(t \midd x)$ is differentiable everywhere, the product integral simplifies to the exponential form $S(t \midd x) = \exp\left\{-\Lambda(t \midd x)\right\}$.
	
	Epidemiological studies are often conducted to learn characteristics of the distribution of time from an initiating event (e.g., disease onset) until a terminating event (e.g., death). Here, we treat the time of the initiating event as $t = 0$, and use the event time $T$ to refer to the time between initiating and terminating events. 
	
	\subsection{Identification}\label{subsec:prospective data structure}
	To start, we consider prospective studies in which individuals who have not yet experienced the event of interest are sampled and followed over time. Ideally, every participant is followed until the event has occurred, but right censoring is essentially inevitable in prospective biomedical studies. Participants who do not experience the event during follow-up are considered right-censored. This may be due to loss to follow-up or to termination of the study.  Let $C \in (0, \infty)$ denote the right censoring time. For each participant in the study, we observe $Y := \min\{T,C\}$, the observed follow-up time, and $\Delta := \I(T \leq C)$, the event indicator. 
	
	Common prospective observational study designs include: (a) the incident cohort --- people who have not experienced the initiating event upon entering the study, and can be followed from the initiating event onward; and (b) the prevalent cohort --- people who experienced the initiating event prior to entering the study. A study sample may also contain both prevalent and incident cases. 
	
	Because prevalent cases have already experienced the initiating event upon study entry, observation of these participants does not begin at $t = 0$. This phenomenon is commonly referred to as delayed entry, and it implies that the event times are observed subject to left truncation. Left truncation induces sampling bias, since individuals with larger event times are more likely to enter the sample. Let $W \in (0,\infty)$ denote the time from the initiating event until entry into the study. Under left truncation, an individual can only enter the study (i.e., be observed) if $W \leq Y$. The observed data for participants in the sample are $O := (X,Y,\Delta,W)$, and the sampling criterion is $W \leq Y$. If a prospective study consists only of incident cases, there is no left truncation. In that special case, $W =0 $ for all participants. 
	
	To identify $\Lambda(\cdot \midd x)$ in the prospective setting, we rely on the following assumption: 
	\begin{itemize}
		\item[] \textit{Assumption A:} $T$ and $(C,W)$ are conditionally independent given $X$. 
	\end{itemize}
	Let $F_{\delta}(y \midd x) := P(Y \leq y \midd  \Delta = \delta, X = x,W\leq Y)$ denote the conditional distribution function of $Y$ among observed participants with $\Delta = \delta$. Let $\pi(x) := P(\Delta = 1 \midd X = x, W \leq Y)$ denote the probability of a random observed individual being uncensored. In addition, define $G_{\delta}(y \midd x) := P(W \leq y \midd  \Delta = \delta,X = x,Y \geq y, W \leq Y)$. We note that these regressions are all functionals of the observed data distribution $P$. As detailed in Section S\ref{sec:calculations} of the Appendix, we then have that $\Lambda(\cdot \midd x)$ can be identified at generic time $t$ by  
	\begin{align}
		\hspace{-0.5cm}\Lambda^{\text{obs}}(t \midd x) := \int_0^t\frac{\pi(x)F_{1}(du \midd x)}{G_{1}(u \midd x)\pi(x)\left\{1 - F_{1}(u^- \midd x)\right\} + G_{0}(u \midd x)\left\{1-\pi(x)\right\}\left\{1 - F_{0}(u^- \midd x)\right\}}\ ,\label{eq:ID1}
	\end{align}
	provided $P^*(W\leq  t\leq C \midd X = x)> 0$.
	When there is no left truncation, the distribution of $W$ is degenerate at 0, so that $G_{1}(u \midd x) = G_{0}(u \midd x) = 1$ for all $u$. In that case, $\Lambda^{\text{obs}}$ is a function only of the conditional distributions of $Y$ given $(\Delta, X)$ and $\Delta$ given  $X$. 
	
	Assumption A can be considered when $C$ is defined for all individuals in the target population. However, in some settings, censoring may only act on enrolled participants. It may then be more appropriate to consider an alternative assumption expressed in terms of residual censoring \citep{Qian2014}. In Section S\ref{sec:calculations}, we show that \eqref{eq:ID1} still holds, and so our proposed estimation strategy is still valid, under one such alternative assumption.
	
	In the retrospective setting, we consider studies in which investigators only sample individuals who have experienced the terminating event prior to the end of the sampling period. For example, in autopsy studies where death is the terminating event, an individual who did not die prior to the end of the study could not enter the sample. This sampling scheme results in right truncation, which, similarly as left truncation, induces sampling bias. In Section S\ref{sec:retrospective} of the Appendix, we show that the above identification results can be directly applied to the retrospective setting by simply considering the time scale to be reversed. 
	
	\subsection{Estimation procedure}\label{sec:estimation}
	
	The results of Section \ref{subsec:prospective data structure} suggest that we can construct an estimator of $\Lambda(\cdot \midd x)$ by estimating a small number of regression functions based on the observed data. This hazard estimator can then be mapped to a survival function estimator via either the product integral or exponential mappings. The regression functions that appear in the identification results constitute either: (i) a conditional probability (specifically, $\pi(x)$), or (ii) a conditional cumulative probability function ($F_1, F_0, G_1$, $G_0$). These regression functions can be estimated using standard machine learning techniques, without requiring any adaptation for the censoring or sampling mechanisms. 
	
	Let $t_\text{max}$ denote the maximum time at which the survival function is to be estimated, and let $t \in (0, t_{\text{max}}]$ be a generic time-point of interest. Below we outline the steps to estimate $S(t \midd x)$. Our proposed procedure is as follows: 
	\setlength{\leftmargini}{1.5em}
	\begin{enumerate}
		
		\item Select an approximation grid: Choose a partition $\mathcal{B}:= \{t_0,t_1\dots,t_\text{max}\}$ of the interval $[0, t_{\text{max}}]$ ($t_0 $ will often be 0).
		
		\item Estimate the cumulative hazard: For each $t_j \in \mathcal{B}$, obtain estimators $F_{1,n}(t_j \midd x)$, $F_{0,n}(t_j \midd x)$, $\pi_n(x)$, $G_{1,n}(t_j \midd x)$, and $G_{0,n}(t_j \midd x)$ of $F_{1}(t_j\midd x)$, $F_{0}(t_j \midd x)$, $\pi(x)$, $G_{1}(t_j \midd x)$, and $G_{0}(t_j \midd x)$ respectively. 
		
		\item Approximate a mapping from the hazard to the survival function:  Let $t_k := \max\{t' \in \mathcal{B}: t' \leq t\}$. Define the estimated differential of the cumulative hazard at $t_i$ as $M_n(t_i, x)$, given by
		\begin{align*}
			\frac{\pi_n( x)\left\{F_{1,n}(t_i \midd x)-F_{1,n}(t_{i-1} \midd x)\right\}}{G_{1,n}(t_i \midd x)\pi_n( x)\left\{1-F_{1,n}(t_{i-1}\midd x)\right\} + G_{0,n}(t_i \midd x)\left\{1-\pi_n(x)\right\}\left\{1-F_{0,n}(t_{i-1}\midd x)\right\}},
		\end{align*}
		where $M_n(t_0, x):= 0$. For the product integral form, approximate the product integral using the product $S_{n,p}(t \midd x):=\prod_{i = 1}^k  \left\{1-M_{n}(t_i,x)\right\}$. For the exponential form, approximate the exponentiated negative cumulative hazard using the Riemann sum approximation
		$S_{n,e}(t \midd x) := \exp\{-\sum_{i=1}^k M_n(t_i,x)\}$.
	\end{enumerate}
	The product integral form of the estimator is the more natural option, since the product integral mapping holds whether $T$ has a discrete, continuous, or mixed distribution. However, in practice, $S_{n,p}$ can have numerical issues, particularly in the right tail of the distribution of $Y$. We discuss this in depth in Section S\ref{sec:additional sims} of the Appendix. 
	
	Approximating the product integral and the cumulative hazard requires choosing an approximation partition $\mathcal{B}$ of the interval $[0, t_{\text{max}}]$. A simple option for this is the set of observed follow-up times $\{Y_{(1)}, Y_{(2)}, \dots, Y_{(n)}\}$, where $Y_{(j)}$ denotes the $j$\textsuperscript{th} order statistic. Alternatively, $\mathcal{B}$ could be set to a fixed number of times between 0 and $t_{\text{max}}$ based on the quantiles of the distribution of $Y$. In large samples, it may be more computationally practical to use a grid of fixed size rather than including every observed follow-up time. 
	
	\subsection{Constructing the constituent regressions}
	
	Our procedure requires estimation of $\pi$, $F_1$, $F_0$, and, when truncation is present, $G_1$ and $G_0$. Estimating the conditional event probability function $\pi(x)$ is a simple binary regression problem, for which there are numerous flexible methods. In practice, we recommend using a boosted classifier, such as boosted trees \citep{Friedman2001}, or an ensemble regression method such as Super Learner \citep{VanDerLaan2007}. The observed follow-up time distribution functions $F_{1}(\cdot \midd x)$ and $F_0(\cdot \midd x)$ are slightly more complicated to estimate. At any fixed time $t$, these can be viewed as a binary regression on the indicator variable $\I(Y \leq t)$. However, we must estimate these distribution functions on the grid $\mathcal{B}$ of times in order to approximate the product- or sum-integral of the hazard function. A simple yet flexible approach is to perform pooled binary regression on a user-specified time grid, which we refer to as the regression grid. We let $\mathcal{C}_1$ and $\mathcal{C}_0$ denote the regression grids used to estimate $F_1$ and $F_0$, respectively. These grids need not be the same as the approximation grid $\mathcal{B}$. Natural choices for $\mathcal{C}_1$ and $\mathcal{C}_0$ would be the observed event times $\mathcal{R}_n := \{Y_i: \Delta_i = 1, i = 1,2,\dots,n\}$ and observed censoring times $\mathcal{S}_n := \{Y_i: \Delta_i = 0, i = 1,2,\dots,n\}$, respectively. We can expect coarser grids (e.g., based on quantiles of the observed event or censoring time distributions) to speed computation at the cost of increased bias. At each time-point $t$ in the grid, the available data are baseline covariates, an indicator outcome variable $\I(Y \leq t)$, and time $t$. These data are pooled across time into a single dataset, which serves as training data for binary regression. This approach differs from local survival stacking in that the risk set at each time-point consists of all participants, and the outcome is cumulative across times. To ensure monotonicity in time, we recommend isotonizing the distribution function estimates using isotonic regression (see for example \citealp{Westling2020}). 
	
	The conditional entry time regression functions $G_{1}(\cdot \midd x)$ and $G_0(\cdot \midd x)$ are similar to conditional distribution functions, although they are each conditioned on being at-risk for an event at time $t$. Similarly as the conditional distribution functions, these functions can be easily estimated using pooled binary regression. Given a time grid, the data at each time-point $t$ consist of all individuals who remain under follow-up at time $t$, along with covariates, time $t$, and the outcome $\I(W \leq t)$. In the simulations, we use the same time grids, based on the observed follow-up times rather than study entry times, to estimate both the $F$ and $G$ regressions. We investigate an alternative approach in Section S\ref{sec:additional sims} of the Appendix.
	
	Because pooled binary regression involves `stacking' datasets across time-points, we refer to the resulting estimation procedure as global survival stacking, which we differentiate from the discrete-time hazard approach of local survival stacking. The global stacking procedure is detailed in Algorithm \ref{alg:cpe}. 
	
	\begin{algorithm}[h!]
		\caption{Global survival stacking}\label{alg:cpe}
		\begin{algorithmic}[1]
			\State Choose grid $\mathcal{B} := \{t_0, t_1, \dots, t_{\text{max}}\}$ for approximation of product- or sum-integral. 
			\State Construct estimator $\pi_n(x)$ of $\pi(x)$ using binary regression. 
			\Algphase{Estimate $F_1$ and $F_0$}
			\For{$\delta \in \{0,1\}$}
			\State Choose grid of time-points
			$\mathcal{C}_\delta := \{t^*_1,t^*_2,\dots,t^*_k\}$ on which to discretize $F_\delta$.
			\State Choose how to include  time in model (continuous, dummy variable, etc.).
			\For{$t^*_j \in \mathcal{C}_\delta$}
			\State Including only participants with $\Delta = \delta$, construct dataset $D_{t^*_j}$ consisting of participant baseline covariates, outcomes $\I(Y \leq t^*_{j})$, and time using chosen basis. 
			\EndFor
			\State Construct full stacked dataset by combining  $\{D_{t_1^*}, D_{t_2^*}, \dots, D_{t_k^*}\}$. 
			\State Fit binary regression or classification algorithm of choice. 
			\State Generate predictions \{$F_{\delta, n}(t_0 \midd x),F_{\delta, n}(t_1 \midd x),\dots,F_{\delta, n}(t_{\text{max}} \midd x)\}$. 
			\EndFor
			\Algphase{Estimate $G_1$ and $G_0$ (if truncation is present)}
			\For{$\delta \in \{0,1\}$}
			\State Choose grid of time-points
			$\mathcal{C}_\delta := \{t^*_1,t^*_2,\dots,t^*_k\}$ on which to discretize $G_\delta$.
			\State Choose how to include  time in model (continuous, dummy variable, etc.).
			\For{$t^*_j \in \mathcal{C}_\delta$}
			\State Including only participants with $\Delta = \delta$ and $Y \geq t^*_j$, construct dataset $D_{t^*_j}$ consisting of participant baseline covariates, outcomes $\I(Y \leq t^*_{j})$, and time using chosen basis. 
			\EndFor
			\State Construct full stacked dataset by combining  $\{D_{t_1^*}, D_{t_2^*}, \dots, D_{t_k^*}\}$. 
			\State Fit binary regression or classification algorithm of choice. 
			\State Generate predictions \{$G_{\delta, n}(t_0 \midd x),G_{\delta, n}(t_1 \midd x),\dots,G_{\delta, n}(t_{\text{max}} \midd x)\}$. 
			\EndFor
			\Algphase{Combine constituent estimators}
			\State Compute $\{M_n(t_0,x),M_n(t_1,x),\dots,M_n(t_{\text{max}},x)\}$, as detailed in Section \ref{sec:estimation}.
			\State Compute $S_{n,p}(t \midd x)$ or $S_{n,e}(t \midd x)$ as detailed in Section \ref{sec:estimation}.
		\end{algorithmic}
	\end{algorithm}
	
	\subsection{Comparison to local survival stacking}\label{sec:comparison}
	Local survival stacking \citep{Polley2011,Craig2021} is a natural alternative to the proposed framework since it allows practitioners to draw upon a wide array of general machine learning techniques. When using local survival stacking, the user must choose how to discretize time. Local survival stacking assumes a discrete survival process, so that the conditional hazard takes the form of a conditional probability that can be estimated for each time-point in the grid. The discretization is usually chosen on the basis of the observed event times $\mathcal{R}_n$. In an illustrative data analysis, \citet{Polley2011} choose 30 time-points based on quantiles of $\mathcal{R}_n$, while \citet{Craig2021} define local survival stacking based on discretizing at each time in $\mathcal{R}_n$. The fineness of the time grid determines the number of events used to estimate the conditional probability of an observed event at each time-point, and we would expect the grid choice to affect performance. The fineness of the time grid may also be relevant for global survival stacking, although we emphasize that the outcome is cumulative over time, meaning that the probability of an outcome at any given time does not shrink as the grid becomes finer. The experiments in Section \ref{sec:simulations} explore the performance of these methods under various grid sizes. In Section S\ref{sec:sim details} of the Appendix, we provide an operational description of local survival stacking.
	
	\section{Results}\label{sec:simulations}
	\subsection{Primary simulation studies}\label{subsec:simulation study}
	We conducted several simulation studies to evaluate the performance of our proposed method. In addition to overall estimation performance compared to other available methods, we aimed to assess the sensitivity of global survival stacking to the choice of approximation time grid $\mathcal{B}$ and regression time grids $(\mathcal{C}_1, \mathcal{C}_0)$. As discussed in Section \ref{sec:comparison}, we expected global survival stacking to be less sensitive to the number of cutpoints in the regression grids compared to local survival stacking due to the fact that the regression outcome is cumulative over time. Furthermore, since including more cutpoints in $\mathcal{B}$ results in a better approximation of the product- or sum-integral, we expected that a finer approximation grid would result in equal or better performance compared to using a coarser grid.
	
	The simulation scenarios are summarized in Table \ref{tab:scenarios}, with Scenarios 1 and 2 described here. We simulated a covariate vector $X := (X_1, X_2, \dots, X_{10})$ of 10 independent components. These components included continuous covariates $X_1, X_2 \sim \text{Uniform}(-1,1)$, discrete covariates $X_3,X_4 \sim \text{Uniform}(\{-1,1\})$, and continuous covariate $X_5 \sim N(0, 1)$. The five additional covariates were independent standard normal noise, i.e., $(X_6,X_7,\dots, X_{10}) \sim \text{MVN}(0, \matr{I}_5)$. Given covariate vector $X = x$, we simulated the censoring time $C$ from a Weibull distribution with shape 1.5 and scale $\lambda_C = \exp\left\{\beta_{0C} + \tfrac{1}{2}\left(x_1 + x_2\right) + \tfrac{1}{5}\left(x_3 + x_4 + x_5\right)\right\}$, where in each simulation setting $\beta_{0C}$ was chosen to give a censoring rate of 25\%. Given covariate vector $X = x$, we independently simulated the event time $T$ to be distributed as $100Z_1$. In the left-skewed scenario, $Z_1$ was a Beta($a(x) + 2, 2$) random variable with $\log a(x) = x_1 + x_2 + x_3 + x_4 + x_5+ x_1x_2 + x_3x_4 + x_1x_5$. In the right-skewed scenario, $Z_1$ was a Beta($2, a(x) + 2$) random variable. Density plots for $T$ given $X$ for 10 random draws from the covariate distribution are given in Section S\ref{sec:sim details}. These distributions do not meet the proportional hazards assumption. Given covariate vector $X = x$, the study entry time variable $W$ was distributed as $100Z_2$, where $Z_2$ was a Beta$\left(1 + \tfrac{1}{2}\I(x_1 > 0), 1 + \tfrac{1}{2}\I(x_1 < 0)\right)$ random variable. In Scenario 2, in which left truncation was present, only observations with $Y \geq W$ were sampled. In Scenario 1, in which there was no truncation, all observations were sampled. The average truncation rates for all simulation settings are given in Section S\ref{sec:sim details}. 
	
	We evaluated performance using Monte Carlo approximations of mean squared error (MSE) at three landmark times and mean integrated squared error (MISE) over the interval $[0,100]$. We computed MSE at landmark times corresponding to the 50\textsuperscript{th}, 75\textsuperscript{th}, and 90\textsuperscript{th} percentiles of observed event times. To calculate the MISE, we computed the MSE at each time on an evenly spaced grid of 1000 points from $t = 0.1$ to $t = 100$, and took a simple average over times. We estimated the performance metrics using a test set of size 1000. The test data were generated without truncation in order to evaluate performance across the marginal distribution of covariates in the target population. 
	
	\begin{table}
		\centering
		\begin{tabular}{ccc}\toprule
			Scenario & Study design & Description \\\midrule
			1 & prospective & right-censored, non-proportional hazards\\
			2 & prospective & left-truncated, right-censored, non-proportional hazards \\
			3 & retrospective & right-truncated, no censoring, non-proportional hazards \\
			4 & prospective & left-truncated, right-censored, proportional hazards \\
			5 & prospective & left-truncated, right-censored, time observed on discrete grid\\
			\bottomrule
		\end{tabular}
		\caption{Simulation scenarios. Results for Scenarios 3 -- 5 are provided in the Appendix.}
		\label{tab:scenarios}
	\end{table} 
	
	We first investigated the sensitivity of global stacking to the number of cutpoints in $\mathcal{B}$, $\mathcal{C}_1$ and  $\mathcal{C}_0$ under Scenario 1. We estimated the binary regressions using a Super Learner consisting of the marginal mean, logistic regression with all pairwise interactions, generalized additive models, multivariate adaptive regression splines, random forests, and gradient-boosted trees. We constructed $\mathcal{B}$, $\mathcal{C}_1$ and $\mathcal{C}_0$ based on the distribution of observed follow-up times, on $\mathcal{R}_n$, and on $\mathcal{S}_n$, respectively. Cutpoints for each grid were evenly spaced on the quantile scale; a full description of the grid choices considered is given in Section S\ref{sec:additional sims} of the Appendix. In Figures \ref{fig:rates A} and \ref{fig:rates B} -- \ref{fig:rates D}, we observe that the performance of global stacking tends to improve as the grids, particularly $\mathcal{B}$, become finer, and eventually reaches a plateau. A relatively coarse grid, e.g., of size $n^{1/2}$, performs nearly as well as finer grids. In the remainder of the experiments, we implemented global stacking with $\mathcal{B}$ set to every observed follow-up time and included several choices for $\mathcal{C}_1$ and $\mathcal{C}_0$, as described below.
	
	\begin{figure}
		\centering
		\includegraphics[width=\linewidth]{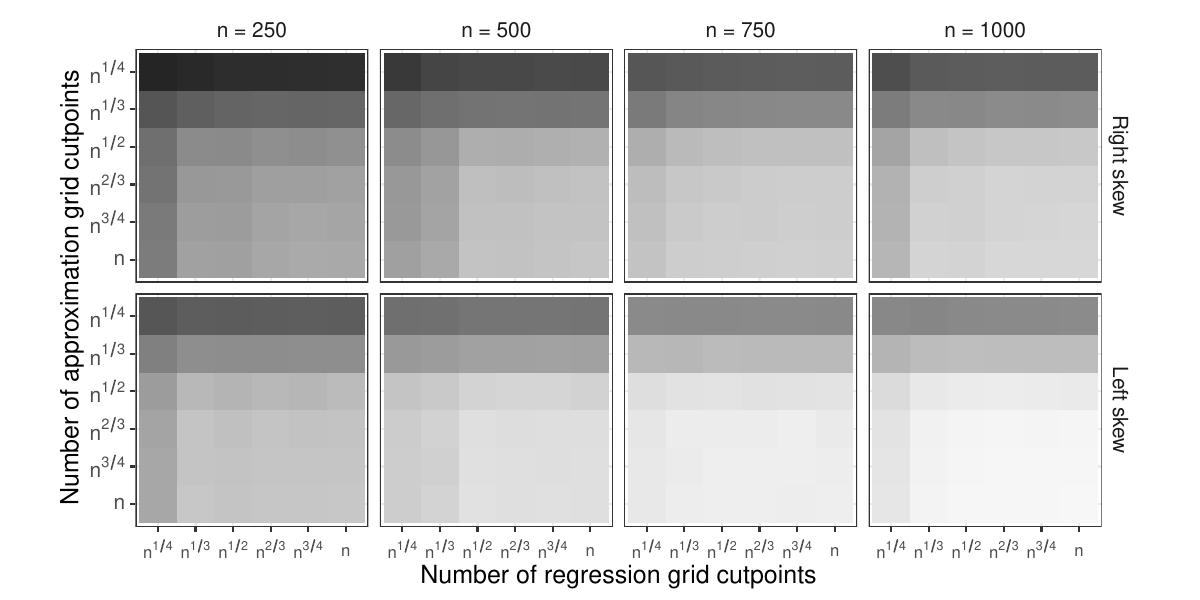}
		\vspace{-0.5cm}
		\caption{Comparison of grid choices for global survival stacking in Scenario 1, with performance measured by MISE. The y-axis represents the number of cutpoints in $\mathcal{B}$, and the x-axis represents the number of cutpoints in $\mathcal{C}_0$ and $\mathcal{C}_1$. Lighter shading indicates lower MISE.}
		\label{fig:rates A}
	\end{figure}
	
	Next, we compared global stacking to several other available methods, which are described in Table \ref{tab:estimators} with full details given in Section S\ref{sec:sim details} of the Appendix. We included local survival stacking, survival Super Learner, random forests (specifically, LTRC conditional inference forests, \citealp{Fu2017}), a linear Cox model, and a generalized additive Cox model \citep{Hastie1986}. For local survival stacking, binary regressions were estimated using a Super Learner with the library detailed above. For global stacking, we considered three options for $\mathcal{C}_1$ and $\mathcal{C}_0$: grids made up of every time in $\mathcal{R}_n$ and $\mathcal{S}_n$, respectively, and grids of 10 or 40 cutpoints evenly spaced on the quantile scales of $\mathcal{R}_n$ and $\mathcal{S}_n$. For local stacking, the same three time grids were included, based on observed event times $\mathcal{R}_n$.
	
	\begin{table}
		\centering
		\begin{tabular}{cccc}\toprule
			Method & Package & Truncation? & Description \\\midrule
			Global surv. stacking & \texttt{survML}&Left, right &Proposed method \\
			Local surv. stacking & \texttt{survML}&Left, right & Discrete-time hazard approach \\
			surv. Super Learner &  \texttt{survSuperLearner}&No & Ensemble survival regression\\
			LTRC forests &  \texttt{LTRCforests}&Left & Conditional inference forest \\
			Linear Cox  & \texttt{survival}& Left & Linear PH model\\
			Gen. additive Cox &\texttt{mgcv} & No & Gen. additive PH model \\ \bottomrule
		\end{tabular}
		\caption{Estimators included in simulation studies. PH indicates proportional hazards.}
		\label{tab:estimators}
	\end{table}

\begin{figure}[h!]
		\centering
		\includegraphics[width=\linewidth]{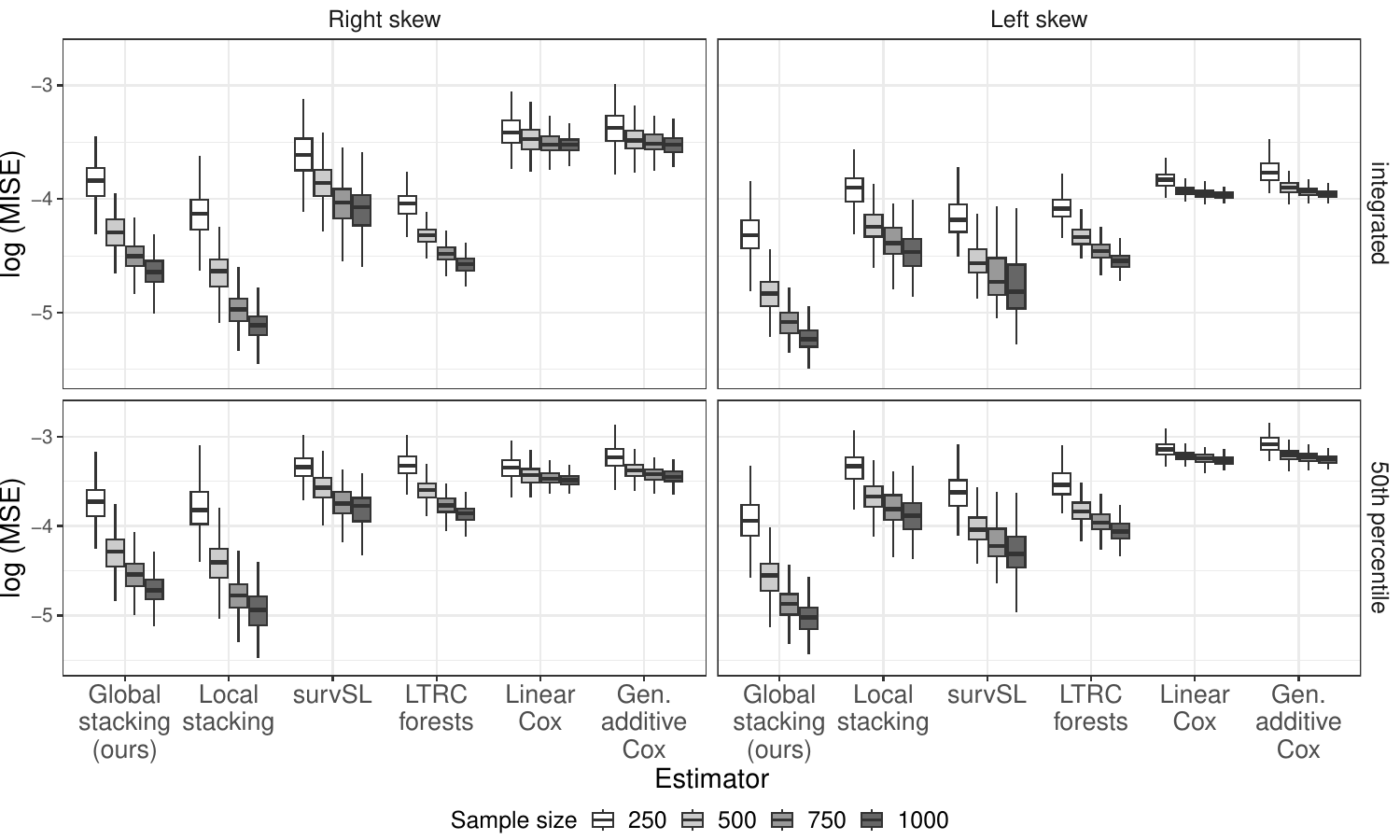}
		\vspace{-0.5cm}
		\caption{Performance of conditional survival estimators with right-censored data (Scenario 1). The methods compared were global survival stacking, local survival stacking, survival Super Learner, LTRC forests, a main-terms linear Cox proportional hazards model with Breslow baseline hazard estimator, and a main-terms generalized additive Cox proportional hazards model with Breslow baseline hazard estimator. Rows correspond to MISE (top) and MSE at the 50th percentile of observed event times (bottom). Each boxplot represents 100 simulation replicates.}
		\label{fig:RC integrated}
	\end{figure}

\begin{figure}[h!]
		\centering
		\includegraphics[width=1\linewidth]{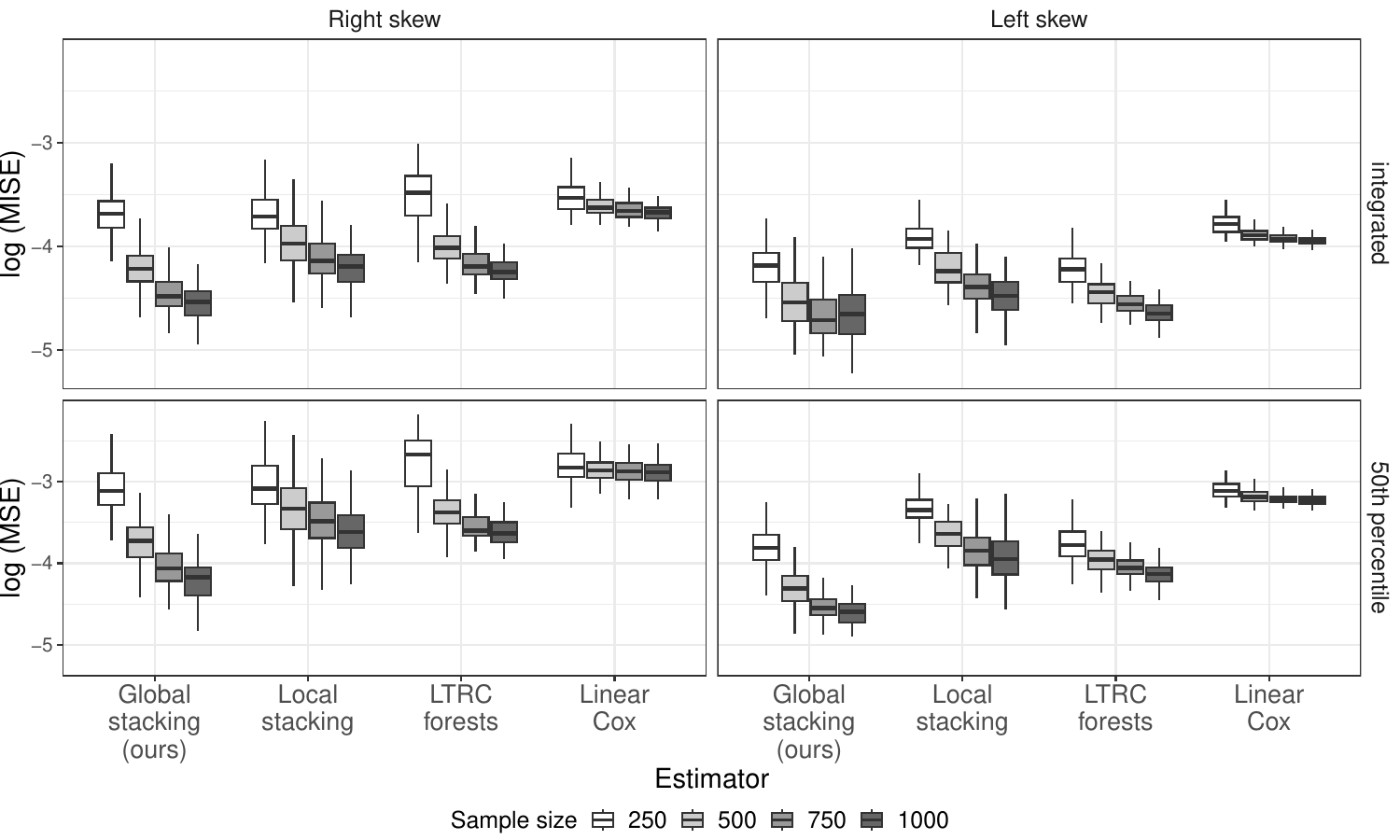}
		\vspace{-0.5cm}
		\caption{Performance of conditional survival estimators with left-truncated, right-censored data (Scenario 2). The methods compared were global survival stacking, local survival stacking, LTRC forests, and a main-terms linear Cox proportional hazards model with Breslow baseline hazard estimator.  Rows correspond to MISE (top) and MSE at the 50th percentile of observed event times (bottom). Each boxplot represents 100 simulation replicates.}
		\label{fig:LTRC integrated}
	\end{figure}

	Figures \ref{fig:RC integrated} and \ref{fig:LTRC integrated} display a subset of results for Scenarios 1 and 2, with global and local stacking implemented using a 40 cutpoint grid. The full results are given in Figures \ref{fig:RC integrated supp} and \ref{fig:LTRC integrated supp}. From Figures \ref{fig:RC integrated} and \ref{fig:LTRC integrated}, we observe that global survival stacking performs well both with and without truncation. The performance of local survival stacking is more variable, but it performs particularly well in the right-skewed setting without truncation. Without truncation, survival Super Learner performs reasonably well, although it is outperformed by global survival stacking in the left-skewed setting and by both global and local survival stacking in the right-skewed setting. The LTRC forests method performs similarly with or without truncation, with performance on par with local stacking in Scenario 2. The Cox model is misspecified, and the performances of both linear and generalized additive Cox models do not improve substantially with sample size. 
	
	From Figures \ref{fig:RC integrated supp} and \ref{fig:LTRC integrated supp}, we see that, as expected, the performance of local survival stacking appears to be more sensitive to grid size choice compared to that of global stacking. Among local stacking implementations, the grid of 40 cutpoints performs the best in general. The 10 cutpoint grid appears too coarse for optimal performance, while the finest grid performs well in the right-skewed settings but poorly in the left-skewed settings.  
	
	\subsection{Additional simulation studies}\label{subsec:computation}
	In Section S\ref{sec:additional sims} of the Appendix, we present additional results. In Scenario 3, we evaluated our procedure under a retrospective study design with right truncation. As in the prospective study design, global stacking demonstrates strong performance, with the finer grids generally outperforming the coarsest grid. In Scenario 4, in which the data were generated from a distribution satisfying the proportional hazards assumption, we found that the correctly specified Cox model yields moderately better performance than the machine learning comparators. Global stacking shows generally good performance, with relatively small differences between different choices of grid size compared with local stacking. While we expect the proportional hazards assumption to rarely hold in practice, predictably, the Cox model would be the preferred method in this situation but with only modest performance loss from more flexible methods. Finally, in Scenario 5, $Y$ and $W$ were observed on a discrete grid of times, rather than in continuous time. When $Y$ and $W$ are observed on a discrete grid of 10 or 20 times, global and local stacking demonstrate similar performance, while global survival stacking performs the best overall when the data are observed on a grid of 50 times. These experiments show that there is relatively little difference between global and local survival stacking when times are observed on a coarse grid, and the advantages of global survival stacking become more pronounced on a finer grid. 
	
	We also conducted a computational benchmarking experiment, finding that local survival stacking, survival Super Learner, and LTRC forests were faster than global survival stacking, and that, unsurprisingly, the computation time required for both global and local survival stacking increases as the number of cutpoints in the grids grows. 
	
	\subsection{Predictive performance on time-to-event datasets}\label{sec:real data performance}
	
	We also evaluated our proposed method on several publicly available datasets with right-censored time-to-event outcomes, which are described in Section S\ref{sec:data details} of the Appendix. We predicted the survival probability at three landmark times, corresponding to the 50\textsuperscript{th}, 75\textsuperscript{th}, and 90\textsuperscript{th} percentiles of observed event times in each dataset, with performance evaluated using the Brier score. To account for censoring, we used the IPCW Brier score given by \citet{Gerds2006}, with Kaplan-Meier censoring weights. For each of the five datasets, we compared global survival stacking, local survival stacking, and LTRC forests. Both global and local stacking were implemented with 40 regression grid cutpoints using the same algorithm library as in the simulations. We also included a na\"ive approach in which, for predicting the survival probability at time $t$, the outcome $\I(Y > t)$ was regressed on $X$. We used the Super Learner, with the same algorithm library as in global and local stacking, to fit this binary regression. We used five-fold cross-validation to estimate the Brier score of each of the methods under consideration. The performance of each method was evaluated relative to the performance of the marginal model constructed without covariates using the Kaplan-Meier estimator (i.e., using the same prediction for every observation in the test set). 
	
	Global survival stacking performs well in all five datasets (Table \ref{tab:public data results}). The na\"ive model typically has relatively poor performance but does slightly outperform the other methods at two landmark times in the SUPPORT dataset, where the censoring rate is zero. This is unsurprising: for prediction at landmark time $t$, observations censored after $t$ provide the same information as uncensored observations. When the censoring rate is higher --- for example, in the METABRIC dataset --- the na\"ive approach is outperformed by the methods that account for censoring. 
	
	\begin{table}[h!]
		\centering
		\begin{tabular}{p{0.11\linewidth}  >{\centering}p{0.11\linewidth} >{\centering}p{0.11\linewidth} >{\centering}p{0.12\linewidth}>{\centering}p{0.12\linewidth} >{\centering}p{0.12\linewidth} >{\centering\arraybackslash}p{0.12\linewidth}} 
			\toprule
			\multirow{3}{*}{Dataset}& \multirow{3}{*}{Quantile}& \multirow{3}{*}{Censoring}&\multicolumn{4}{c}{Performance relative to KM} \\
			\cmidrule{4-7}&&&Global stacking &Local stacking&LTRC forests&\multirow{2}{*}{Na\"ive}\\\midrule
			FLCHAIN &50\textsuperscript{th}&0.07&\textbf{0.749}&0.751&0.775&0.757\\
			&75\textsuperscript{th}&0.19&\textbf{0.686}&0.689&0.708&0.694\\
			&90\textsuperscript{th}&0.31&\textbf{0.647}&0.657&0.672&0.662\\
			GBSG &50\textsuperscript{th}&0.03&\textbf{0.855}&0.886&0.876&0.889 \\
			&75\textsuperscript{th}&0.07&\textbf{0.825}&0.841&0.858&0.971\\
			&90\textsuperscript{th}&0.17&\textbf{0.838}&0.850&0.861&1.125\\
			METABRIC &50\textsuperscript{th}&0.07&\textbf{0.891}&0.908&0.914&0.913\\
			&75\textsuperscript{th}&0.19&\textbf{0.885}&0.887&0.901&0.981\\
			&90\textsuperscript{th}&0.30&0.873&\textbf{0.870}&0.878&1.043\\
			NWTCO &50\textsuperscript{th}&0.02&\textbf{0.861}&\textbf{0.861}&0.919&0.863 \\
			&75\textsuperscript{th}&0.04&\textbf{0.867}&0.875&0.907&0.895\\
			&90\textsuperscript{th}&0.11&\textbf{0.866}&0.873&0.902&0.993\\
			SUPPORT &50\textsuperscript{th}&0.00&0.930&0.952&0.948&\textbf{0.927}\\ 
			&75\textsuperscript{th}&0.00&\textbf{0.909}&0.923&0.920&\textbf{0.909}\\
			&90\textsuperscript{th}&0.08&\textbf{0.879}&0.892&0.886&0.903\\\bottomrule
		\end{tabular}
		\caption{Predictive performance of candidate methods on publicly available survival datasets. The performance metric is the Brier score standardized by the Brier score of the Kaplan-Meier (KM) estimator (i.e., predicting survival probability without using covariate information). The Brier score was evaluated at three landmark times corresponding to the 50\textsuperscript{th}, 75\textsuperscript{th}, and 90\textsuperscript{th} percentiles of observed event times. Lower values are preferred. Boldface font indicates the best performance for each dataset and landmark time. The methods compared were global survival stacking (our proposed method), local survival stacking, LTRC forests, and a na\"ive binary regression approach ignoring censoring.}
		\label{tab:public data results}
	\end{table}
	
	\subsection{Assessing risk of HIV infection in the STEP trial}\label{sec:step}
	
	Between December 2004 and March 2007, 3,000 HIV-negative individuals were enrolled in the STEP study (HVTN 502/Merck 023), a randomized, placebo-controlled phase 2b trial that tested the efficacy of a candidate HIV vaccine to prevent acquisition of HIV-1 infection. The vaccine contains an adenovirus serotype 5 (Ad5) vector that expresses subtype B HIV-1 \textit{gag/pol/nef} proteins. Participants were at high risk of HIV-1 acquisition. Participants were unblinded in October 2007 after the prespecified monitoring boundary for efficacy futility was crossed at the first interim analysis \citep{Buchbinder2008}. Data analyses suggested an increased risk of HIV-1 infection among vaccine recipients versus placebo recipients, particularly among participants who were uncircumcised or had neutralizing antibodies against the Ad5 vector at enrollment (``baseline Ad5 titer”).
	
	To assess the risk of HIV-1 infection conditional on circumcision status and baseline Ad5 titer, we estimated the conditional survival function of the time-to-infection-diagnosis variable within randomized treatment arms at landmark times of one year and two years of follow-up, corresponding to approximately 60\% and 10\% of participants still at-risk in each treatment arm. We limited our analyses to the 1,836 participants with male sex assigned at birth in the modified intention-to-treat cohort, which included all vaccinated participants except those diagnosed as HIV-1 positive on or before the day 1 visit. At one year of follow-up, 41 participants in the vaccine arm (4.6\%) and 27 participants in the placebo arm (3.0\%) had been diagnosed with HIV-1; at two years, 51 participants in the vaccine arm (5.7\%) and 35 participants in the placebo arm (3.9\%) had been diagnosed. We implemented global survival stacking using Super Learner with the same algorithm library as in the simulations, using regression grids of 40 cutpoints with five-fold cross-validation for tuning. For comparison, we also fit a Cox model including the circumcision/baseline Ad5 titer interaction and estimated the baseline cumulative hazard function using the Breslow estimator. Both models were fit separately in the two treatment arms. Baseline Ad5 titer was log-transformed (using the natural logarithm), and titers under the assay detection limit of 18 were treated as equal to 18 for analysis \citep{Duerr2012}. We calculated the risk difference conditional on circumcision status and baseline Ad5 titer by taking the difference of the estimated conditional survival functions in the two treatment arms at each landmark time. Using global survival stacking, we also computed representative survival curves for individuals in each treatment arm, circumcised and uncircumcised, at log baseline Ad5 titer values of 3, 5, and 7. 
	
\begin{figure}[h!]
		\centering
		\includegraphics[width=\linewidth]{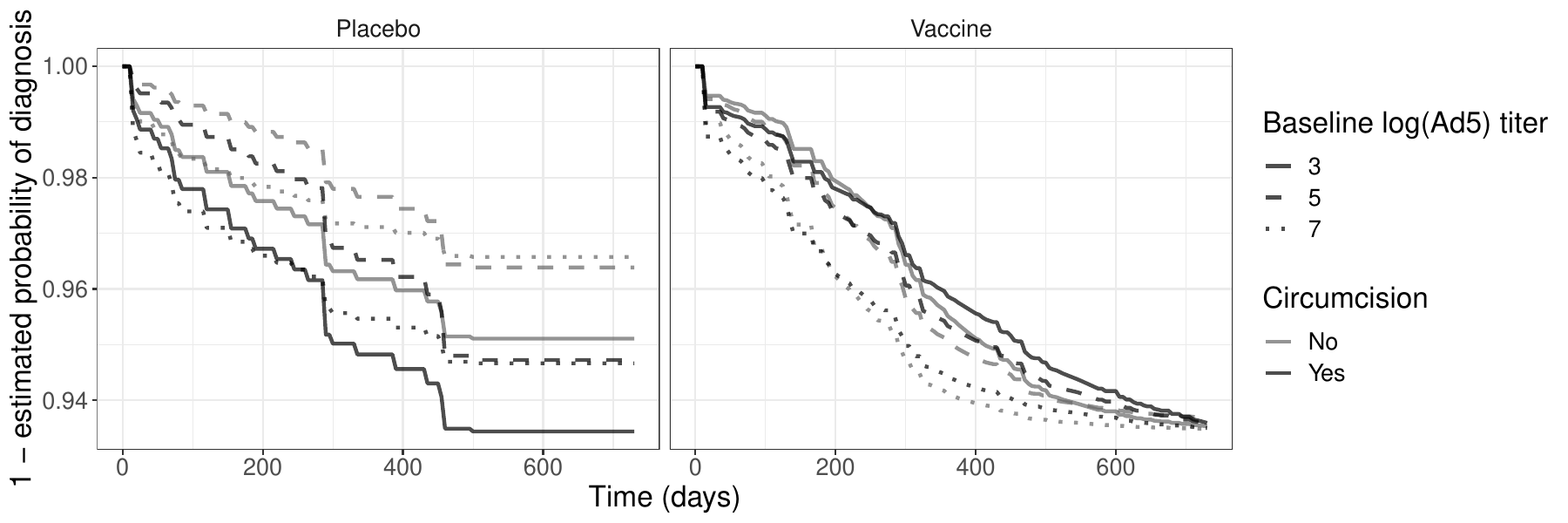}
		\vspace{-0.25cm}
		\caption{Estimated survival curves for time to HIV-1 diagnosis in the STEP study. The curves were estimated separately in each treatment arm, conditional on baseline Ad5 titer and circumcision status.}
		\label{fig:step curves}
	\end{figure}

\begin{figure}[h!]
		\centering
		\includegraphics[width=\linewidth]{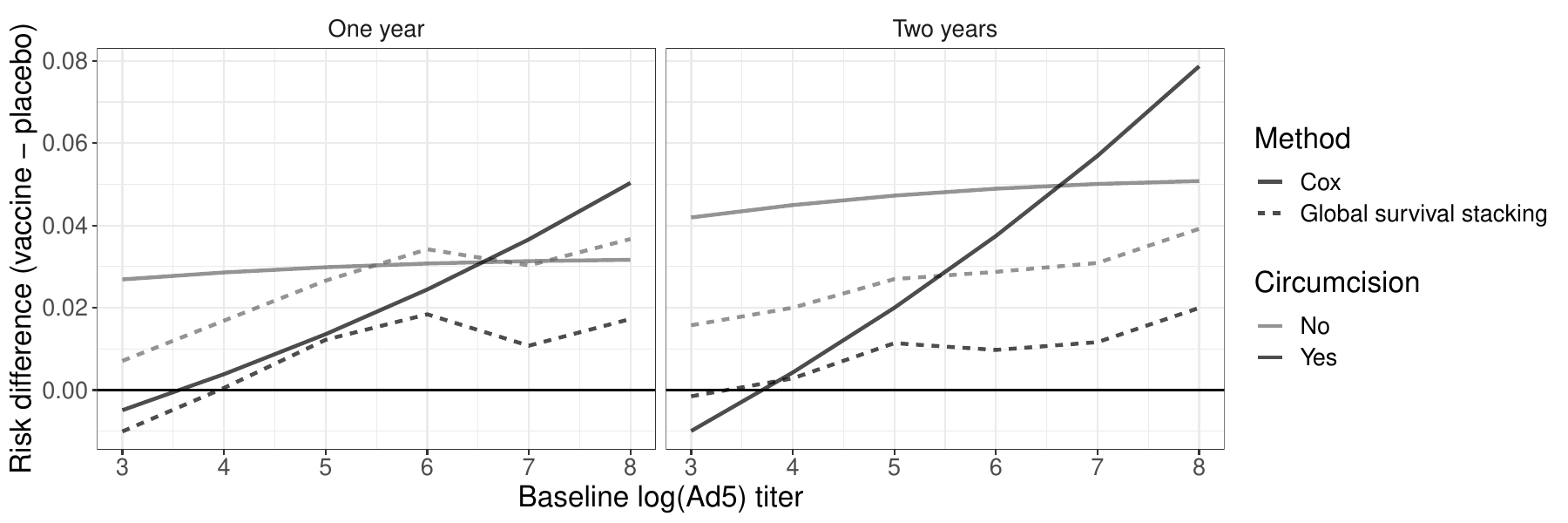}
		\vspace{-0.25cm}
		\caption{Estimated risk difference (vaccine - placebo) of HIV-1 infection diagnosis in the STEP study conditional on baseline Ad5 titer and circumcision status at one year and two years of follow-up. The estimators compared were the Cox model with first-order interaction, and global survival stacking.}
		\label{fig:step}
	\end{figure}

	The estimated survival curves (Figure \ref{fig:step curves}) show that, in the vaccine group, the probability of HIV-1 diagnosis through day 730 tends to be higher for individuals with higher baseline Ad5 titers. The probability of HIV-1 diagnosis was higher in the vaccine arm than the placebo arm, as estimated by both global stacking and the Cox model, except at low baseline Ad5 titers among circumcised participants (Figure \ref{fig:step}). The estimated excess risk in the vaccine arm tends to increase with baseline Ad5 titer and is generally higher among uncircumcised participants, although the Cox model fit suggests that circumcised participants may have slightly larger excess risk at high baseline Ad5 titers. Overall, these results agree with the original analysis in \citet{Duerr2012}, which did not explicitly account for right censoring.
	
	\section{Discussion}\label{sec:discussion}
	In this article, we proposed a framework for estimating a conditional survival function in both prospective and retrospective settings using flexible machine learning tools. This framework, which we call global survival stacking, relies on an identification of the hazard function in terms of observable regressions that can be estimated using standard methods for binary outcomes, without the need to explicitly account for censoring or truncation. Similarly as with local survival stacking, our approach recasts conditional survival function estimation as a statistical learning task that does not require specially tailored survival analysis tools. These methods not only enable practitioners to take advantage of the myriad machine learning methods currently available, but also to harness the improved performance of new methods as they are developed. Numerical experiments show that global survival stacking works well across a variety of settings, performing on par with or better than competing methods when the proportional hazards assumption fails to hold. The performance of global stacking tends to improve as the number of cutpoints used in $\mathcal{B}$, $\mathcal{C}_1$ and $\mathcal{C}_0$ increases. 
	
	Both global and local survival stacking can be computationally expensive, particularly when time grids with large numbers of cutpoints are used. Global survival stacking requires fitting multiple regressions on data sets that are generally larger than those that arise in local survival stacking. Based on the performance of global survival stacking, we recommend using time grids that are as fine as computational resources and time allow. Using a large number of cutpoints in $\mathcal{B}$ has a relatively small computational cost compared to doing so for $\mathcal{C}_0$ or $\mathcal{C}_1$. If an analysis is only performed once on a dataset of modest size, using regression grids of every observed follow-up time may be reasonable. However, in our experiments there was little gain for the computational cost, and global stacking suffered minimal decreases in performance from using relatively coarse grids. In practice, of course, the computational resources required for using any ensemble regression method will depend on which algorithms are included in the library. A theoretical analysis of the impact of grid choice could help provide additional guidance for the use of global stacking in practice.
	
	Because our method involves estimating regression functions within strata defined by the event indicator, we can use the same procedure to obtain an estimate of the conditional censoring distribution, simply replacing $\pi_n(x)$ with $1 - \pi_n(x)$ and $F_{1,n}(t \midd x)$ with $F_{0,n}(t \midd x)$ in the numerator of $M_n(t,x)$ in Section \ref{sec:estimation}. While any conditional survival function estimation algorithm can be repurposed by reversing the roles of $T$ and $C$, our approach requires no refitting, resulting in greater computational efficiency when both distributions are desired. 
	
	Unlike right censoring, interval censoring is a type of data coarsening that remains unaddressed by many survival function estimators. Interval-censored event times are known only to lie in a particular interval, rather than being observed exactly. Data that are truly subject to interval censoring (e.g., data from biomedical studies with periodic follow-up) are often treated as subject only to right censoring. Whether the machine learning methods presented here can be adapted to handle interval censored data remains an open question.
	
	\section{Acknowledgments}
	The authors thank the study participants and investigators of the STEP HVTN 502/Merck 023 trial conducted by the HIV Vaccine Trials Network.  The authors also thank Alex Luedtke for his insightful comments. Research reported in this publication was supported by National Institute Of Allergy And Infectious Diseases grants UM1-AI068635 and R37-AI029168, National Heart, Lung, and Blood Institute grant R01-HL137808, and National Science Foundation Graduate Research Fellowship Program under Grant No. DGE-2140004. 
	
	\newpage 
	\bibliography{refs} 

@article{Fu2017,
  title={Survival trees for left-truncated and right-censored data, with application to time-varying covariate data},
  author={Fu, Wei and Simonoff, Jeffrey S},
  journal={Biostatistics},
  volume={18},
  number={2},
  pages={352--369},
  year={2017},
  publisher={Oxford University Press}
}

@article{Knaus1995,
  title={The SUPPORT prognostic model: Objective estimates of survival for seriously ill hospitalized adults},
  author={Knaus, William A and Harrell, Frank E and Lynn, Joanne and Goldman, Lee and Phillips, Russell S and Connors, Alfred F and Dawson, Neal V and Fulkerson, William J and Califf, Robert M and Desbiens, Norman and others},
  journal={Annals of internal medicine},
  volume={122},
  number={3},
  pages={191--203},
  year={1995},
  publisher={American College of Physicians}
}

@article{Dangio1976,
   author = {Giulio J. D'Angio and Audrey E. Evans and Norman Breslow and Bruce Beckwith and Harry Bishop and Polly Feigl and Willard Goodwin and Lucian L. Leape and Lucius F. Sinks and Wataru Sutow and Melvin Tefft and James Wolff},
   doi = {10.1002/1097-0142(197608)38:2<633::AID-CNCR2820380203>3.0.CO;2-S},
   issn = {10970142},
   issue = {2},
   journal = {Cancer},
   pages = {633-646},
   pmid = {184912},
   title = {The treatment of Wilms' tumor. Results of the national Wilms' tumor study},
   volume = {38},
   year = {1976},
}

@article{Schumacher1994,
  title={Randomized 2 x 2 trial evaluating hormonal treatment and the duration of chemotherapy in node-positive breast cancer patients. German Breast Cancer Study Group.},
  author={Schumacher, M and Bastert, G and Bojar, H and H{\"u}bner, K and Olschewski, M and Sauerbrei, W and Schmoor, C and Beyerle, C and Neumann, RL and Rauschecker, HF},
  journal={Journal of Clinical Oncology},
  volume={12},
  number={10},
  pages={2086--2093},
  year={1994}
}

@Manual{survival-package,
    title = {A Package for Survival Analysis in R},
    author = {Terry M Therneau},
    year = {2022},
    note = {R package version 3.3-1},
    url = {https://CRAN.R-project.org/package=survival},
  }

@article{Kyle2006,
  title={Prevalence of monoclonal gammopathy of undetermined significance},
  author={Kyle, Robert A and Therneau, Terry M and Rajkumar, S Vincent and Larson, Dirk R and Plevak, Matthew F and Offord, Janice R and Dispenzieri, Angela and Katzmann, Jerry A and Melton III, L Joseph},
  journal={New England Journal of Medicine},
  volume={354},
  number={13},
  pages={1362--1369},
  year={2006},
  publisher={Mass Medical Soc}
}

@article{Curtis2012,
  title={The genomic and transcriptomic architecture of 2,000 breast tumours reveals novel subgroups},
  author={Curtis, Christina and Shah, Sohrab P and Chin, Suet-Feung and Turashvili, Gulisa and Rueda, Oscar M and Dunning, Mark J and Speed, Doug and Lynch, Andy G and Samarajiwa, Shamith and Yuan, Yinyin and others},
  journal={Nature},
  volume={486},
  number={7403},
  pages={346--352},
  year={2012},
  publisher={Nature Publishing Group UK London}
}

@article{Gerds2006,
   author = {Thomas A. Gerds and Martin Schumacher},
   doi = {10.1002/bimj.200610301},
   issn = {03233847},
   issue = {6},
   journal = {Biometrical Journal},
   pages = {1029-1040},
   pmid = {17240660},
   title = {Consistent estimation of the expected {B}rier score in general survival models with right-censored event times},
   volume = {48},
   year = {2006},
}

@techreport{Beran1981,
   author = {Rudolf Beran},
   issn = {0303-6898},
   title = {Non-parametric regression with censored survival time data},
   year = {1981},
   institution = {University of California, Berkeley},
}

@article{Breslow1972,
   author = {Norman E. Breslow},
   issue = {2},
   journal = {Journal of the Royal Statistical Society. Series B (Statistical Methodology)},
   pages = {216-217},
   title = {Discussion of the paper by {D}.{R}. {Cox}.},
   volume = {34},
   year = {1972},
}

@article{Buchbinder2008,
   author = {Susan P. Buchbinder and Devan V. Mehrotra and Ann Duerr and Daniel W. Fitzgerald and Robin Mogg and David Li and Peter B. Gilbert and Javier R. Lama and Michael Marmor and Carlos del Rio and M. Juliana McElrath and Danilo R. Casimiro and Keith M. Gottesdiener and Jeffrey A. Chodakewitz and Lawrence Corey and Michael N. Robertson},
   doi = {10.1016/S0140-6736(08)61591-3},
   issn = {01406736},
   journal = {The Lancet},
   pages = {1881-1893},
   pmid = {19012954},
   title = {Efficacy assessment of a cell-mediated immunity {H}{I}{V}-1 vaccine (the {S}tep Study): a double-blind, randomised, placebo-controlled, test-of-concept trial},
   volume = {372},
   year = {2008},
}

@article{Cox1972,
   author = {D. R. Cox},
   isbn = {00359246},
   issn = {00359246},
   issue = {2},
   journal = {Journal of the Royal Statistical Society. Series B (Methodological)},
   pages = {187-220},
   pmid = {2985181},
   title = {Regression Models and Life-Tables},
   volume = {34},
   year = {1972},
}

@article{Craig2021,
   author = {Erin Craig and Chenyang Zhong and Robert Tibshirani},
   journal = {arXiv:2107.13480},
   title = {Survival stacking: casting survival analysis as a classification problem},
   url = {http://arxiv.org/abs/2107.13480},
   year = {2021},
}

@article{Diaz2019,
   author = {Iván Díaz},
   doi = {10.1002/sim.8156},
   issn = {10970258},
   issue = {15},
   journal = {Statistics in Medicine},
   pages = {2735-2748},
   pmid = {30950107},
   publisher = {John Wiley and Sons Ltd},
   title = {Statistical inference for data-adaptive doubly robust estimators with survival outcomes},
   volume = {38},
   year = {2019},
}

@article{Duerr2012,
   author = {Ann Duerr and Yunda Huang and Susan Buchbinder and Robert W. Coombs and Jorge Sanchez and Carlos Del Rio and Martin Casapia and Steven Santiago and Peter Gilbert and Lawrence Corey and Michael N. Robertson},
   doi = {10.1093/infdis/jis342},
   issn = {00221899},
   issue = {2},
   journal = {Journal of Infectious Diseases},
   pages = {258-266},
   pmid = {22561365},
   title = {Extended follow-up confirms early vaccine-enhanced risk of {H}{I}{V} acquisition and demonstrates waning effect over time among participants in a randomized trial of recombinant adenovirus HIV vaccine ({S}tep Study)},
   volume = {206},
   year = {2012},
}

@article{Fleming1984,
   author = {Thomas R Fleming and David P Harrington},
   issue = {20},
   journal = {Communications in Statistics: Theory and Methods},
   pages = {2469-2486},
   title = {Nonparametric estimation of the survival distribution in censored data},
   volume = {13},
   year = {1984},
}

@article{Gill1990,
   author = {Richard D. Gill and Soren Johansen},
   issue = {4},
   journal = {The Annals of Statistics},
   pages = {1501-1555},
   title = {A Survey of Product-Integration with a View Toward Application in Survival Analysis},
   volume = {18},
   year = {1990},
}

@article{Harrell1982,
   author = {Frank E. Harrell and Robert M. Califf and David B. Pryor and Kerry L. Lee and Robert A. Rosati},
   doi = {10.1001/jama.1982.03320430047030},
   issn = {15383598},
   issue = {18},
   journal = {JAMA: The Journal of the American Medical Association},
   pages = {2543-2546},
   pmid = {7069920},
   title = {Evaluating the Yield of Medical Tests},
   volume = {247},
   year = {1982},
}

@article{Hastie1986,
   author = {Trevor Hastie and Robert Tibshirani},
   issue = {3},
   journal = {Statistical Science},
   pages = {297-318},
   title = {Generalized Additive Models},
   volume = {1},
   year = {1986},
}

@article{Hothorn2006,
   author = {Hothorn, T. and Bühlmann, P. and Dudoit, S. and van der Laan, M. J.},
   doi = {10.1093/biostatistics/kxj011},
   issn = {14654644},
   issue = {3},
   journal = {Biostatistics},
   pages = {355-373},
   pmid = {16344280},
   title = {Survival ensembles},
   volume = {7},
   year = {2006},
}

@article{Ishwaran2008,
   author = {Hemant Ishwaran and Udaya B. Kogalur and Eugene H. Blackstone and Michael S. Lauer},
   doi = {10.1214/08-AOAS169},
   issn = {19326157},
   issue = {3},
   journal = {Annals of Applied Statistics},
   pages = {841-860},
   title = {Random survival forests},
   volume = {2},
   year = {2008},
}

@article{Kaplan1958,
   author = {E.L. Kaplan and Paul Meier},
   isbn = {01621459},
   issn = {01621459},
   issue = {282},
   journal = {Journal of the American Statistical Association},
   pages = {457-481},
   pmid = {14670573},
   title = {Nonparametric Estimation from Incomplete Observations},
   volume = {53},
   year = {1958},
}

@article{Katzman2018,
   author = {Jared L. Katzman and Uri Shaham and Alexander Cloninger and Jonathan Bates and Tingting Jiang and Yuval Kluger},
   doi = {10.1186/s12874-018-0482-1},
   issn = {14712288},
   issue = {24},
   journal = {BMC Medical Research Methodology},
   pages = {1-12},
   pmid = {29482517},
   title = {DeepSurv: Personalized treatment recommender system using a {C}ox proportional hazards deep neural network},
   volume = {18},
   year = {2018},
}

@inbook{Polley2011,
    title={Super Learning for Right-Censored Data},
   author = {Eric C. Polley and Mark J. van der Laan},
   pages = {249-258},
   publisher = {Springer},
   booktitle = {Targeted Learning: Causal Inference for Observational Data},
   year = {2011},
   city = {New York},
}

@article{Tarkhan2022,
  title={An online framework for survival analysis: reframing Cox proportional hazards model for large data sets and neural networks},
  author={Tarkhan, Aliasghar and Simon, Noah},
  journal={Biostatistics},
  year={2022}
}

@article{vanderlaan2007,
  title={Super learner},
  author={van der Laan, MJ and Polley, EC and Hubbard, AE},
  journal={Statistical Applications in Genetics and Molecular Biology},
  volume={6},
  number={1},
  pages={Online Article 25},
  year={2007}
}

@book{VanderLaan2003,
   author = {Mark J. van der Laan and James M. Robins},
   city = {New York},
   publisher = {Springer},
   title = {Unified Methods for Censored Longitudinal Data and Causality},
   year = {2003},
}

@article{Wei1992,
   author = {L. J. Wei},
   issue = {14-15},
   journal = {Statistics in Medicine},
   pages = {1871-1879},
   title = {The accelerated failure time model: A useful alternative to the {C}ox regression model in survival analysis.},
   volume = {11},
   year = {1992},
}

@article{Westling2023,
  title={Inference for treatment-specific survival curves using machine learning},
  author={Westling, Ted and Luedtke, Alex and Gilbert, Peter B and Carone, Marco},
  journal={Journal of the American Statistical Association},
  number={just-accepted},
  pages={1--26},
  year={2023},
  publisher={Taylor \& Francis}
}

@article{Yu2011,
   author = {Chun Nam Yu and Russell Greiner and Hsiu Chin Lin and Vickie Baracos},
   isbn = {9781618395993},
   journal = {Advances in Neural Information Processing Systems},
   pages = {1845-1853},
   title = {Learning patient-specific cancer survival distributions as a sequence of dependent regressors},
   volume = {24},
   year = {2011},
}

@article{Zeng2007,
   author = {Donglin Zeng and D. Y. Lin},
   doi = {10.1198/016214507000001085},
   issn = {01621459},
   issue = {480},
   journal = {Journal of the American Statistical Association},
   pages = {1387-1396},
   title = {Efficient estimation for the accelerated failure time model},
   volume = {102},
   year = {2007},
}

@article{Zhou2022,
   author = {Xingyu Zhou and Wen Su and Changyu Liu and Yuling Jiao and Xingqiu Zhao and Jian Huang},
   journal = {arXiv:2205.09633},
   title = {Deep Generative Survival Analysis: Nonparametric Estimation of Conditional Survival Function},
   url = {http://arxiv.org/abs/2205.09633},
   year = {2022},
}

@article{Friedman2001,
   author = {Jerome H Friedman},
   issue = {5},
   journal = {The Annals of Statistics},
   pages = {1189-1232},
   title = {Greedy Function Approximation: A Gradient Boosting Machine},
   volume = {29},
   year = {2001},
}

@article{Westling2020,
   author = {Ted Westling and Mark J. van der Laan and Marco Carone},
   doi = {10.1214/20-EJS1740},
   issn = {19357524},
   issue = {2},
   journal = {Electronic Journal of Statistics},
   pages = {3032-3069},
   publisher = {Institute of Mathematical Statistics},
   title = {Correcting an estimator of a multivariate monotone function with isotonic regression},
   volume = {14},
   year = {2020},
}

@article{Lin2013,
   author = {Yuanyuan Lin and Kani Chen},
   doi = {10.1093/biomet/ass073},
   issn = {00063444},
   issue = {2},
   journal = {Biometrika},
   pages = {525-530},
   title = {Efficient estimation of the censored linear regression model},
   volume = {100},
   year = {2013},
}

@article{Qian2014,
   author = {Jing Qian and Rebecca A. Betensky},
   doi = {10.1016/j.spl.2013.12.016},
   issn = {01677152},
   issue = {1},
   journal = {Statistics and Probability Letters},
   pages = {12-17},
   title = {Assumptions regarding right censoring in the presence of left truncation},
   volume = {87},
   year = {2014},
}
	\newpage
	
	\titleformat{\section}{\normalfont\scshape\bfseries}{S\arabic{section}.}{1em}{}
	\titleformat{\subsection}{\normalfont\bfseries}{S\arabic{section}.\arabic{subsection}}{1em}{}
	\renewcommand{\thefigure}{S\arabic{figure}}
	\setcounter{figure}{0}
	\renewcommand{\thetable}{S\arabic{table}}
	\setcounter{table}{0}
	\setcounter{section}{0}
	
	\section{Retrospective sampling}\label{sec:retrospective}
	
	As in the prospective setting, we let $W$ be the study entry time. We do not consider censoring in this setting, so $T$ is observed for all participants. For notational consistency, we set $C = 0$ for all participants and define $Y := \max\{T,C\}$ and $\Delta := \I(T \geq C) = 1$. This implies that $Y = T$ for all participants, i.e., the observed follow-up times are equal to the event times. Under right truncation, an individual is sampled if $Y \leq W$. The observed data are $O := (X,Y,\Delta,W)$, and the sampling criterion is $W \geq Y$.
	
	Identification of $S(t \midd x)$ in the retrospective setting follows from the prospective identification. Let $\tau$ denote a user-specified real number, and  define the random variables $\bar{T} := \tau - T$, $\bar{C} := \tau - C$, $\bar{Y} := \tau - Y$, $\bar{W} := \tau - W$, and $\bar{\Delta} := \I(\bar{T} \leq \bar{C}) = 1$. In practice, we set $\tau$ as the maximum study entry time $W$, so that $\bar{T}$, $\bar{C}$, $\bar{Y}$ and $\bar{W}$ are non-negative. If $T$ has bounded support, the upper bound of that support would be another natural choice for $\tau$. (In principle, $\tau$ could be any real number, including 0, in which case the transformed data could take negative values. For the sake of applying our prospective results to the retrospective setting, we assume the transformed data are nonnegative.) We suppose that Assumption A holds. 
	
	We note that $\bar{T}$ is subject to conditionally independent left truncation by $\bar{W}$. Denoting by $\bar{\Lambda}(t \midd x)$ and $\bar{S}(t \midd x)$ the conditional cumulative hazard and survival functions of $\bar{T}$ given $X$ at $t$, we can directly use the prospective setting results to identity $\bar{\Lambda}(\cdot \midd x)$ at generic time point $t$ by
	\begin{align*}
		\int_0^{t}\frac{\bar{\pi}(x)\bar{F}_{1}(du \midd x)}{\bar{G}_{1}(u \midd x)\bar{\pi}(x)\left\{1 - \bar{F}_{1}(u^- \midd x)\right\} + \bar{G}_{0}(u \midd x)\left\{1-\bar{\pi}(x)\right\}\left\{1 - \bar{F}_{0}(u^- \midd x)\right\}}\ ,
	\end{align*}
	where $\bar{F}_1, \bar{F}_0$, $\bar{G}_1$, and $\bar{G}_0$ are defined analogously as in the prospective setting, and $\bar{\pi}(x):= P(\bar{\Delta} = 1 \midd X = x)$. Because there is no censoring in this setting, $\bar{\pi}(x) = 1$, and the above identification can be written in the form
	\begin{align*}
		\bar{\Lambda}^{\text{obs}}(t \midd x) := \int_0^{t}\left[\bar{G}_{1}(u \midd x)\left\{1 - \bar{F}_{1}(u^- \midd x)\right\}\right]^{-1}\bar{F}_{1}(du \midd x)\ .
	\end{align*}
	Finally, we note that $S(t \midd x)$ can be written as $1- \bar{S}(\tau - t \midd x)$,
	and so it suffices to use the above identification of $\bar{\Lambda}(\cdot \midd x)$ in order to estimate $S(\cdot \midd x)$. This result demonstrates that estimating the conditional hazard of $T$ given  $X$ under right truncation can be accomplished by simply estimating the conditional hazard of $\tau - T$ given  $X$ under left truncation. 
	
	In the retrospective setting, estimation proceeds by simply (i) transforming the data to reverse time, taking $\bar{Y}_i = \tau - Y_i$, $\bar{\Delta}_i = 1$, $\bar{W}_i = \tau - W_i$, and $\bar{t}  = \tau - t$; (ii) following Steps 1, 2, and 3 in Section \ref{sec:estimation} of the main text to produce an estimate $\bar{S}_n(\bar{t} \midd x)$ of $\bar{S}(\bar{t} \midd x)$, with $\bar{S}_n$ being either the product integral or exponential form; and (iii) computing $S_n(t \midd x) = 1 - \bar{S}_{n}(\bar{t} \midd x)$. 
	
	\section{Details of identification result}\label{sec:calculations}
	Let $F_{T, C, W}$ and $F_{C,W}$ denote the conditional distribution functions of $(T,C,W)$ given $X$ and $(C,W)$ given  $X$, respectively. We begin by using standard probability rules to write
	\begin{align*}
		\pi(x)F_1(u \midd x) \ &=\  P(\Delta = 1\midd X = x, W \leq Y)P(Y \leq u\midd \Delta = 1, X = x, W \leq Y)\\
		&=\  \frac{P(\Delta =1, Y \leq u, W \leq Y \midd X = x)}{P(W \leq Y \midd X = x)}\\
		&=\  \frac{P(T \leq C, T \leq u, W \leq T \midd X = x)}{P(W \leq Y \midd X = x)}\\
		&=\  \frac{\iiint \I(t \leq u, c \geq t, w \leq t)F_{T,C,W}(dt, dc, dw \midd x)}{P(W \leq Y \midd X = x)}\\
		&\stackrel{\text{(a)}}{=}\  \frac{ \iiint\I(t \leq u, c \geq t, w \leq t)F_{C,W}(dc, dw \midd x)F(dt\midd x)}{P(W \leq Y \midd X = x)}\\
		&=\  \frac{ \int_0^uP(C \geq t, W \leq t\midd X = x)F(dt \midd x)}{P(W \leq Y \midd X = x)}\  ,
	\end{align*}
	where (a) follows from Assumption A. The differential of this function with respect to $u$ is 
	\begin{align*}
		\pi(x)F_1(du \midd x) = \frac{P(W \leq u, C \geq u  \midd X = x)F(du \midd x)}{P(W \leq Y \midd X = x)}\ .
	\end{align*}
	The denominator of $\Lambda^{\text{obs}}$ is
	\begin{align*}
		&G_1(u \midd x)\pi(x)\left\{1 - F_1(u^- \midd x)\right\} +G_0(u \midd x)\left\{1-\pi(x)\right\}\left\{1 - F_0(u^- \midd x)\right\} \\
		&=\ P(W \leq u, Y \geq u \midd W \leq Y, X = x)\ ,
	\end{align*}
	where we have applied the law of total probability. Continuing from this expression we have
	\begin{align*}
		P(W \leq u, Y \geq u \midd W \leq Y, X = x) \ &=\  \frac{P(W \leq u \leq Y, W \leq Y \midd X = x)}{P(W \leq Y \midd X = x)}\\
		&=\  \frac{P(W \leq u \leq Y\midd X = x)}{P(W \leq Y \midd X = x)}\\
		&=\  \frac{P(W \leq u, C \geq u, T \geq u \midd  X = x)}{P(W \leq Y \midd X = x)}\\
		&\stackrel{\text{(b)}}{=}\  \frac{P(W \leq u, C \geq u \midd X = x)S(u^{-} \midd  x)}{P(W \leq Y \midd X = x)}\ , 
	\end{align*}
	where (b) follows from Assumption A. Combining this with the numerator, we have
	\begin{align*}
		&\Lambda^{\text{obs}}(t \midd x) \ =\  \int_0^t\frac{\pi(x)F_{1}(du \midd x)}{G_{1}(u \midd x)\pi(x)\left\{1 - F_1(u^- \midd x)\right\} +G_{0}(u \midd x)\left\{1-\pi(x)\right\}\left\{1 - F_0(u^- \midd x)\right\}}\\
		&\hspace{0.5cm}=\  \int_0^t \left. \left(\frac{P(W \leq u, C \geq u  \midd X = x)}{P(W \leq Y \midd X = x)}\right) \middle/ \left( \frac{P(W \leq u, C \geq u \midd X = x)S(u^{-} \midd x)}{P(W \leq Y \midd X = x)}\right)F(du \midd x)\right.\\
		&\hspace{0.5cm}=\  \int_0^t\frac{F(du\midd x)}{S(u^{-} \midd x)}\\
		&\hspace{0.5cm}=\  \Lambda(t \midd x) \ .
	\end{align*}
	
	\section{Identification under alternative assumption}\label{sec:quasi}
	In this section, we consider an alternative identifying assumption for use in contexts in which censoring can only occur in individuals who satisfy the sampling criterion. Assumption B is given in three parts as:
	\begin{itemize}
		\item[] \textit{Assumption B1:} $W < C$ almost surely;
		\item[] \textit{Assumption B2:} $T$ and $W$ are conditionally independent given $X$;
		\item[] \textit{Assumption B3:} $T$ and $C$ are conditionally independent given $(X,W)$ and $W\leq T$.
	\end{itemize}
	Let $F_W$ denote the conditional distribution function of $W$ given $X$. Let $H_{T,C,W}$ and $H_W$ denote respectively the conditional distribution functions of $(T,C,W)$ and $W$ given both $X$ and $W \leq T$. Let  $H_{T,C \midd W}$, $H_{C\midd W}$, and $H_{T \midd W}$ denote respectively the conditional distribution functions of $(T,C)$, $C$, and $T$ given both $( W, X)$ and $W \leq T$.
	
	We note that Assumption (B2) allows us to write
	\begin{align}
		\label{eq:eq3}
		H_{T \midd W}(t \midd w,x) \ &=\ P(T \leq t \midd X = x, W = w, W \leq T)\nonumber\\
		&=\ P(T \leq t \midd X = x, W = w, T \geq w)\nonumber\\
		&=\ \frac{P(T \leq t, T \geq w \midd X = x, W = w)}{P(T \geq w \midd X = x, W = w)}\nonumber\\
		&=\ \frac{\I(w \leq t)P(w \leq T \leq t \midd X = x, W = w)}{P(T \geq w \midd X = x, W = w)}\nonumber\\
		&\stackrel{\text{(a)}}{=}\ \frac{\I(w \leq t)P(w \leq T \leq t \midd X = x)}{P(T \geq w \midd X = x)}\ ,
	\end{align}
	where (a) follows from Assumption B2. We then use standard probability rules to write
	\begin{align*}
		\pi(x)F_1(u \midd x) \ &=\  P(\Delta = 1\midd X = x, W \leq Y)P(Y \leq u\midd \Delta = 1, X = x, W \leq Y)\\
		&=\  P(\Delta =1, Y \leq u \midd W \leq Y, X = x)\\
		&\stackrel{\text{(b)}}{=}\ P(\Delta =1, Y \leq u \midd W \leq T, X = x)\\
		&=\  P(T \leq C, T \leq u \midd W \leq T, X = x)\\
		&=\  \iiint \I(t \leq u, c \geq t)H_{T,C,W}(dt, dc, dw \midd x)\\
		&=\ \iiint \I(t \leq u, c \geq t)H_{T,C|W}(dt, dc \midd w, x)H_W(dw \midd x)\\
		&\stackrel{\text{(c)}}{=}\ \iiint \I(t \leq u, c \geq t)H_{C|W}(dc \midd w, x)H_{T|W}(dt \midd w, x)H_W(dw \midd x)\\
		&\stackrel{\text{(d)}}{=}\ \iiint \frac{\I(t \leq u, c \geq t, w \leq t)H_{C|W}(dc \midd w, x)F(dt \midd x) H_W(dw \midd x)}{P(T \geq w \midd X = x)}\ ,
	\end{align*}
	where (b) follows from Assumption B1, (c) from Assumption B3, and (d) from equation \eqref{eq:eq3}. The differential of this function with respect to $u$ is 
	\begin{align*}
		\pi(x)F_1(du \midd x) = \iint \frac{\I(c \geq u, w \leq u)H_{C|W}(dc \midd w, x) H_W(dw \midd x)F(du \midd x)}{P(T \geq w \midd X = x)}\ .
	\end{align*}
	Let $R(u,x) := \iint \frac{\I(c \geq u, w \leq u)H_{C|W}(dc \midd w, x) H_W(dw \midd x)}{P(T \geq w \midd X = x)}$. As in Section S\ref{sec:calculations}, the denominator of $\Lambda^{\text{obs}}$ can be written as
	\begin{align*}
		P(W \leq u, &Y \geq u \midd W \leq Y, X = x) \ \stackrel{\text{(e)}}{=}\ P(W \leq u, Y \geq u \midd W \leq T, X = x)\\
		&=\ P(W \leq u, C \geq u, T \geq u \midd W \leq T, X = x)\\
		&= \iiint \I(t \geq u, c \geq u, w \leq u)H_{T,C,W}(dt, dc, dw \midd x)\\
		&=\ \iiint \I(t \geq u, c \geq u, w \leq u)H_{T,C|W}(dt, dc \midd w, x)H_W(dw \midd x)\\
		&\stackrel{\text{(f)}}{=}\ \iiint \I(t \geq u, c \geq u, w \leq u)H_{C|W}(dc \midd w, x)H_{T|W}(dt \midd w, x)H_W(dw \midd x)\\
		&\stackrel{\text{(g)}}{=}\ \iiint \frac{\I(t \geq u, c \geq u, w \leq u, w \leq t)H_{C|W}(dc \midd w, x)F_T(dt \midd x) H_W(dw \midd x)}{P(T \geq w \midd X = x)}\\
		&=\ \iiint \frac{\I(t \geq u, c \geq u, w \leq u)H_{C|W}(dc \midd w, x)F_T(dt \midd x) H_W(dw \midd x)}{P(T \geq w \midd X = x)}\\
		&=\ S(u^- \midd x)R(u,x)\ ,
	\end{align*}
	where (e) follows from Assumption B1, (f) from Assumption B3, and (g) from equation \eqref{eq:eq3}. Combining this with the numerator, we find, as claimed, that
	\begin{align*}
		\Lambda^{\text{obs}}(t \midd x) \ &=\  \int_0^t\frac{\pi(x)F_{1}(du \midd x)}{G_{1}(u \midd x)\pi(x)\left\{1 - F_1(u^- \midd x)\right\} +G_{0}(u \midd x)\left\{1-\pi(x)\right\}\left\{1 - F_0(u^- \midd x)\right\}}\\
		&=\  \int_0^t \frac{R(u,x)}{R(u,x)S(u^{-} \midd x)} F(du \midd x)\ =\   \int_0^t\frac{F(du\midd x)}{S(u^{-} \midd x)} \ =\  \Lambda(t \midd x) \ .
	\end{align*}

	\section{Simulation details}\label{sec:sim details}
	
	\subsection{Additional details on estimators and data-generating mechanism}
	
	Here, we describe the estimators included in the simulation studies. The \texttt{R} package implementation is given in parentheses. 
	
	\begin{enumerate}
		\item Global survival stacking (\texttt{survML}): We estimated $F_1, F_0, G_1$, and $G_0$ using pooled binary regression with Super Learner, as implemented in the \texttt{SuperLearner} software package. The algorithm library consisted of the marginal mean, logistic regression with all pairwise interactions, generalized additive models, multivariate adaptive regression splines, random forests, and gradient-boosted trees. We estimated $\pi(x)$  using the same Super Learner library. We used five-fold cross-validation and the built-in nonnegative least-squares method to determine
		the optimal convex combination of these algorithms. In all experiments, $\mathcal{B}$ was based on the observed follow-up times, $\mathcal{C}_1$ was based on $\mathcal{R}_n:= \{Y_i: \Delta_i = 1\}$, and $\mathcal{C}_0$ was based on $\mathcal{S}_n := \{Y_i: \Delta_i = 0\}$. Grids were evenly spaced on the quantile scale. In the experiments in Section \ref{subsec:grid rates}, we set the number of cutpoints in each grid to take values in $\{n^{1/4}, n^{1/3}, n^{1/2}, n^{2/3}, n^{3/4}, n\}$. In the remainder of the experiments, we set $\mathcal{B}$ to the set of all observed follow-up times and considered three regression time grids: grids made up of every time in $\mathcal{R}_n$ and $\mathcal{S}_n$, and grids of 40 or 10 cutpoints. We refer to these grids as ``fine," ``medium," and ``coarse." We used the exponential form $S_{n,e}(t \midd x)$. The predictions across times in the approximation grid were isotonized using the pool adjacent violators algorithm, as implemented in the \texttt{Iso} software package. Table \ref{tab:SL algorithms} details the Super Learner library used for estimating binary regressions in global and local survival stacking. 
		
		\item Local survival stacking (\texttt{survML}): We used Super Learner as the binary classifier in local survival stacking, using the same algorithm library as for global survival stacking. Tuning was performed in the same manner as described above, and the same time grids were included, based on observed event times $\mathcal{R}_n$. Algorithm \ref{alg:stack prospective} details the procedure to construct the local survival stacking algorithm. 
		
		\item Survival Super Learner (\texttt{survSuperLearner}): We used the same library of algorithms for both the censoring and event time distributions, including the marginal Kaplan-Meier estimator, the Cox proportional hazards model with Breslow baseline hazard estimator, exponential regression, Weibull regression, log-logistic regression, a generalized additive proportional hazards model, and random survival forest. We did not evaluate this method in any settings with truncation since it is not designed to handle truncation. Table \ref{tab:survSL algorithms} details the Super Learner library used for \texttt{survSuperLearner}.
		
		\item Random forests (\texttt{LTRCforests}): We used conditional inference forests for left-truncated, right-censored data. We set the \texttt{mtry} parameter equal to the square root of the number of predictors, rounded up. 
		
		\item Linear Cox proportional hazards regression (\texttt{survival}): We used a main-terms linear Cox proportional hazards model with Breslow baseline hazard estimator. 
		
		\item Generalized additive Cox proportional hazards regression (\texttt{mgcv}):  We used a main-terms generalized additive Cox proportional hazards model with Breslow baseline hazard estimator. 
	\end{enumerate} 
	
	Below, we include additional information on the simulation data-generating mechanisms. 
	
\begin{figure}
		\begin{center}
			\includegraphics[width=\linewidth]{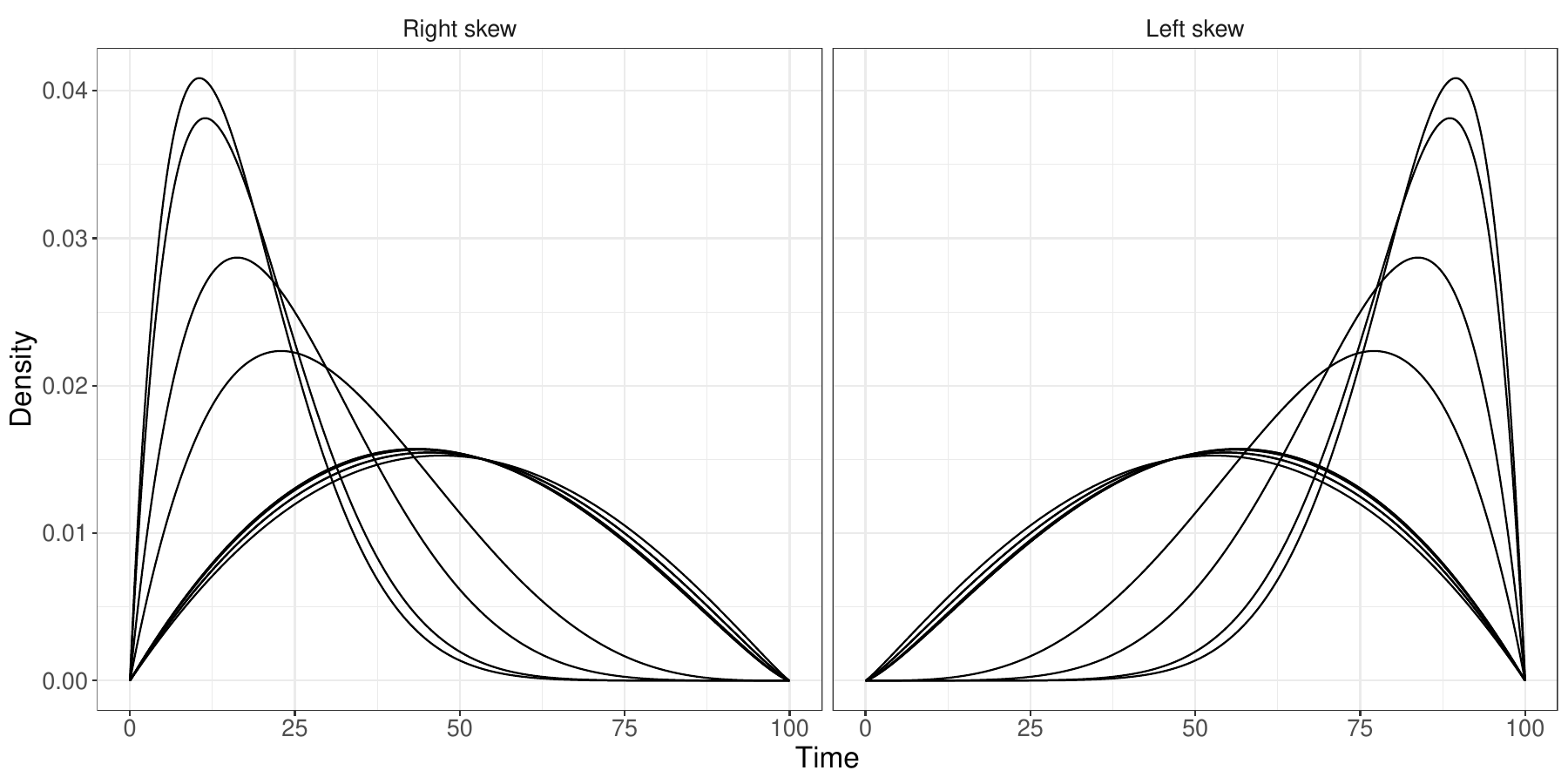}
		\end{center}
		\vspace{-0.5cm}
		\caption{Example densities for the time-to-event variable $T$ under the two data-generating mechanisms used in simulations. Each plot shows the conditional density of $T$ given $X$ for ten random draws from the distribution of $X$.}
		\label{fig:example densities}
	\end{figure}

	\begin{itemize}
		
		\item[-] \textbf{Figure \ref{fig:example densities}:} Example densities for the time-to-event variable $T$ under the two data-generating mechanisms used in simulations.
		
		\item[-] \textbf{Table \ref{tab:cens trunc}:} Average truncation rates in numerical experiments. 
	\end{itemize}
	
	\renewcommand\thealgorithm{S\arabic{algorithm}}
	\begin{algorithm}
		\caption{Local survival stacking}\label{alg:stack prospective}
		\begin{algorithmic}[1]
			\State Choose grid of time-points
			$\mathcal{C} := \{t^*_1,t_2^*,\dots,t^*_k\}$ on which to discretize. Set $t^*_{k+1} = \infty$.
			\State Choose how to include  time in model (continuous, dummy variable, etc.).
			\For{$t^*_j \in \mathcal{C}$}
			\State Including only participants with $Y \geq t_j^*$ and $W \leq t_j^*$, construct dataset $D_{t^*_j}$ consisting of participant baseline covariates, outcomes $\I(t^*_j \leq Y < t^*_{j+1})$, and time using chosen basis. 
			\EndFor
			\State Construct full stacked dataset by combining $\{D_{t_1^*}, D_{t_2^*}, \dots, D_{t_k^*}\}$. 
			\State Fit binary regression or classification algorithm of choice. 
			\State Generate hazard predictions $\{\lambda_n(t^*_1 \midd x), \lambda_n(t^*_2 \midd x), \dots, \lambda_n(t^*_k \midd x) \}$ from fitted model.
			\State Compute estimate $S_n(t \midd x) = \prod_{t^*_j \in \mathcal{C}: t^*_j \leq t} \{1-\lambda_n(t_j^* \midd x)\}$. 
		\end{algorithmic}
	\end{algorithm}
	
	\begin{table}
		\centering 
		\begin{tabular}{p{0.25\linewidth}p{0.35\linewidth}p{0.3\linewidth} }\toprule 
			Algorithm name & Algorithm description & Tuning parameters\\ \midrule 
			\texttt{SL.mean} & Marginal mean&NA\\\cmidrule{1-3}
			\texttt{SL.glm.interaction} & Logistic regression with pairwise interactions&NA\\ \cmidrule{1-3}
			\texttt{SL.gam} & Generalized additive model&default\\\cmidrule{1-3}
			\texttt{SL.earth} & Multivariate adaptive regression splines&default\\\cmidrule{1-3}
			\texttt{SL.ranger} & Random forest&default\\\cmidrule{1-3}
			\texttt{SL.xgboost} & Gradient-boosted trees&\texttt{ntrees} $\in \{250, 500, 1000\}$\\
			&&\texttt{max\_depth} $\in \{1, 2\}$\\\bottomrule
		\end{tabular}
		\caption{Algorithms included in the Super Learner for global and local survival stacking. All tuning parameters besides those for \texttt{SL.xgboost} were set to default values. In particular, \texttt{gam} was implemented with \texttt{degree = 2}; \texttt{earth} with \texttt{degree = 2, penalty = 3, nk =} number of predictors plus 1, \texttt{endspan = 0}, \texttt{minspan = 0}; and \texttt{ranger} with \texttt{num.trees = 500, mtry =} the square root of the number of predictors, \texttt{min.node.size = 1, sample.fraction = 1} with replacement. For \texttt{SL.xgboost}, \texttt{shrinkage} was set to 0.01, \texttt{minobspernode} was set to 1, and each combination of \texttt{ntrees} and \texttt{max\_depth} was included in the Super Learner library.}
		\label{tab:SL algorithms}
	\end{table} 
	
	\begin{table}
		\centering
		\begin{tabular}{p{0.25\linewidth}  p{0.65\linewidth}}\toprule
			Algorithm name & Algorithm description \\ \midrule
			\texttt{survSL.km} & Kaplan-Meier estimator\\	\cmidrule{1-2}
			\texttt{survSL.expreg} & Survival regression assuming event and censoring times follow an exponential distribution conditional on covariates\\ 	\cmidrule{1-2}
			\texttt{survSL.weibreg} & Survival regression assuming event and censoring times follow a Weibull distribution conditional on covariates\\	\cmidrule{1-2}
			\texttt{survSL.loglogreg} & Survival regression assuming event and censoring times follow a log-logistic distribution conditional on covariates\\	\cmidrule{1-2}
			\texttt{survSL.gam} & Main-terms generalized additive Cox proportional hazards estimator as implemented in the \texttt{mgcv} package\\	\cmidrule{1-2}
			\texttt{survSL.coxph} & Main-terms Cox proportional hazards estimator with Breslow baseline cumulative hazard\\	\cmidrule{1-2}
			\texttt{survSL.rfsrc} & Random survival forest as implemented in the \texttt{randomForestSRC} package\\\bottomrule
		\end{tabular}
		\caption{Algorithms included in the survival Super Learner. All tuning parameters were set to default values. In particular, \texttt{gam} was implemented with \texttt{degree = 1}; and \texttt{rfsrc} with \texttt{ntree = 500}, \texttt{mtry =} the square root of the number of predictors, \texttt{nodesize = 15}, \texttt{splitrule = "logrank"}, \texttt{sampsize = 1} with replacement.}
		\label{tab:survSL algorithms}
	\end{table}

	\begin{table}
		\centering 
		\begin{tabular}{l c c   c} 
			\toprule 
			Study design & Skew &  Setting& Truncation rate\\ \midrule 
			Prospective & Right & Non-proportional hazards &70\%\\
			&& Proportional hazards &66\%\\
			&& Discrete & 70\% \\
			&Left & Non-proportional hazards&46\%\\
			&& Proportional hazards &51\%\\
			&& Discrete & 46\% \\
			Retrospective&Right&Non-proportional hazards &35\%\\
			&Left & Non-proportional hazards&65\%\\\bottomrule
		\end{tabular}
		\caption{Average truncation rates across simulations.}
		\label{tab:cens trunc}
	\end{table}

	\section{Additional numerical results}\label{sec:additional sims}

	\subsection{Comparison of choices for $\mathcal{B}$, $\mathcal{C}_1$ and $\mathcal{C}_0$}\label{subsec:grid rates}
	
\begin{figure}
		\centering
		\includegraphics[width=\linewidth]{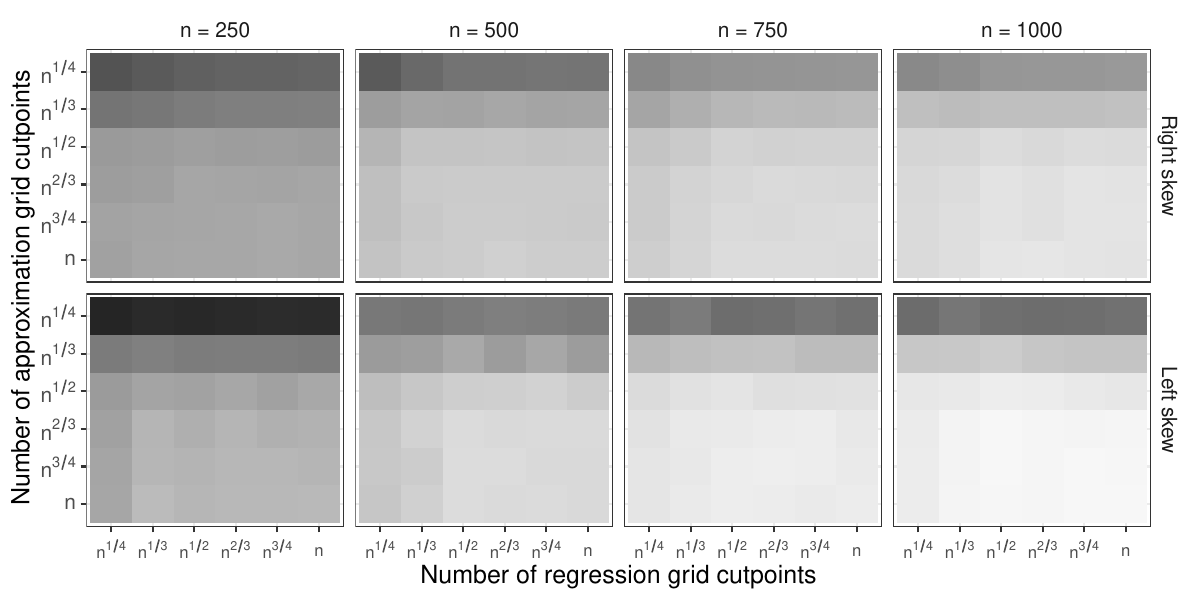}
		\vspace{-0.5cm}
		\caption{Comparison of grid choices for global survival stacking in Scenario 1, with performance measured by MSE at the 50\textsuperscript{th} percentile of observed event times. The y-axis represents the number of cutpoints in $\mathcal{B}$, and the x-axis represents the number of cutpoints in $\mathcal{C}_0$ and $\mathcal{C}_1$. Lighter shading indicates lower MSE.}
		\label{fig:rates B}
	\end{figure}

\begin{figure}
		\centering
		\includegraphics[width=\linewidth]{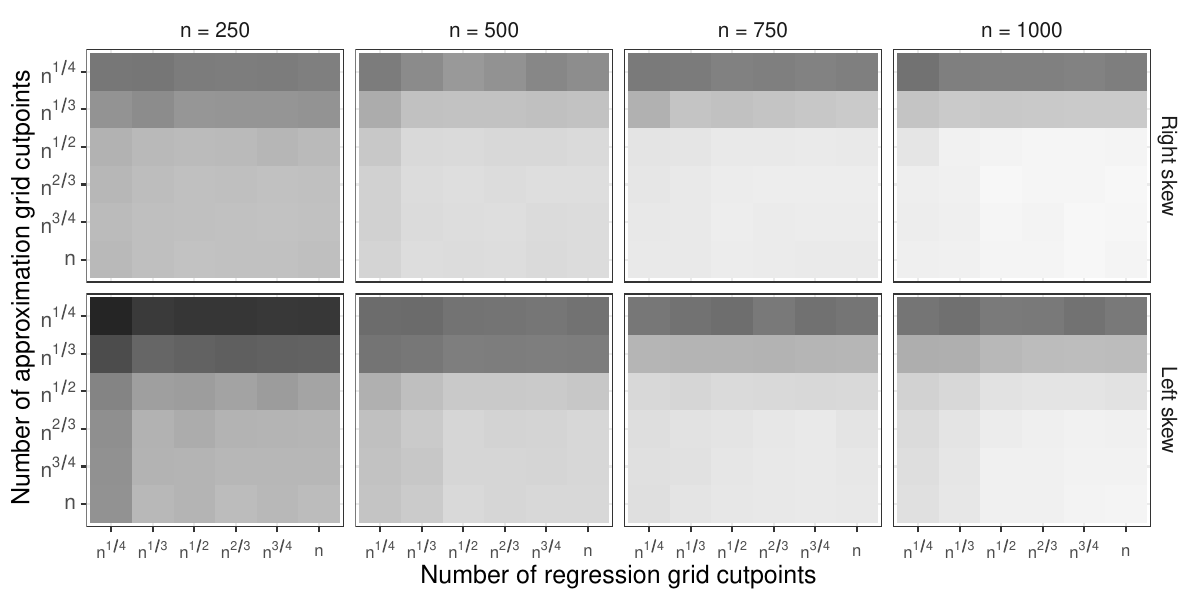}
		\vspace{-0.5cm}
		\caption{Comparison of grid choices for global survival stacking in Scenario 1, with performance measured by MSE at the 75\textsuperscript{th} percentile of observed event times. The y-axis represents the number of cutpoints in $\mathcal{B}$, and the x-axis represents the number of cutpoints in $\mathcal{C}_0$ and $\mathcal{C}_1$. Lighter shading indicates lower MSE.}
		\label{fig:rates C}
	\end{figure}

\begin{figure}
		\centering
		\includegraphics[width=\linewidth]{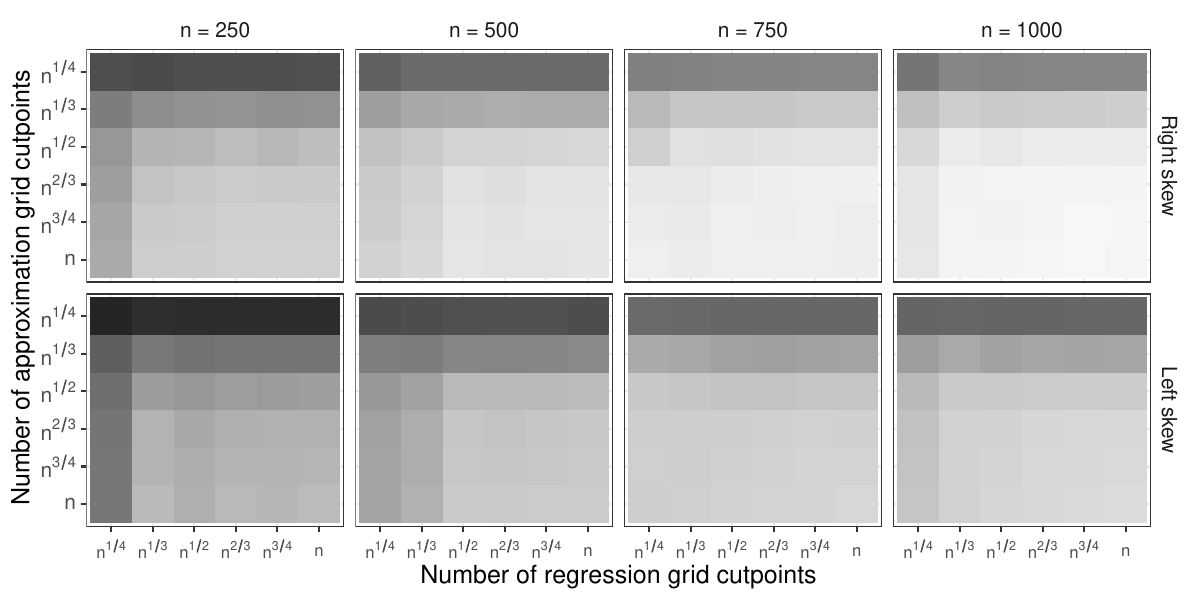}
		\vspace{-0.5cm}
		\caption{Comparison of grid choices for global survival stacking in Scenario 1, with performance measured by MSE at the 90\textsuperscript{th} percentile of observed event times. The y-axis represents the number of cutpoints in $\mathcal{B}$, and the x-axis represents the number of cutpoints in $\mathcal{C}_0$ and $\mathcal{C}_1$. Lighter shading indicates lower MSE.}
		\label{fig:rates D}
	\end{figure}

	This simulation was performed under Scenario 1, as described in Section \ref{subsec:simulation study} of the main text. Figures \ref{fig:rates A} and \ref{fig:rates B} -- \ref{fig:rates D} display the results for MISE, MSE at the 50\textsuperscript{th} percentile of observed event times, MSE at the 75\textsuperscript{th} percentile of observed event times, and MSE at the 90\textsuperscript{th} percentile of observed event times. By all performance metrics, we see that increasing the number of cutpoints in both the approximation and regression grids tends to improve performance, although the improvements are minimal when the number of cutpoints increases beyond $n^{1/2}$. Table \ref{tab:rate runtimes} shows the computation time for each of the grid choices. Unsurprisingly, there is a significantly larger cost to increasing the number of cutpoints in $\mathcal{C}_1$ and $\mathcal{C}_0$ compared to $\mathcal{B}$.
	
	\begin{table}
		\centering
		\begin{tabular}{ccrrrrrr}
			\toprule
			\multirow{2}{*}{Training sample size} & \multirow{2}{*}{\# cutpoints in $\mathcal{C}$ }&\multicolumn{6}{c}{\# cutpoints in $\mathcal{B}$ }\\
			\cmidrule{3-8}&  & $n^{1/4}$ & $n^{1/3}$ & $n^{1/2}$ & $n^{2/3}$ & $n^{3/4}$ & $n$ \\ 
			\hline
			n = 250 & $n^{1/4}$ & 53 & 54 & 57 & 73 & 86 & 177 \\ 
			& $n^{1/3}$ & 53 & 58 & 62 & 76 & 89 & 181 \\ 
			& $n^{1/2}$ & 66 & 69 & 76 & 90 & 105 & 202 \\ 
			& $n^{2/3}$ & 101 & 98 & 106 & 123 & 136 & 242 \\ 
			& $n^{3/4}$ & 127 & 129 & 134 & 152 & 166 & 272 \\ 
			& $n$ & 268 & 279 & 278 & 294 & 318 & 428 \\ 
			n = 500 & $n^{1/4}$ & 68 & 67 & 80 & 100 & 128 & 340 \\ 
			& $n^{1/3}$ & 73 & 74 & 85 & 110 & 135 & 355 \\ 
			& $n^{1/2}$ & 112 & 118 & 129 & 154 & 184 & 433 \\ 
			& $n^{2/3}$ & 238 & 233 & 250 & 273 & 311 & 600 \\ 
			& $n^{3/4}$ & 384 & 367 & 385 & 417 & 460 & 723 \\ 
			& $n$ & 1281 & 1336 & 1296 & 1344 & 1384 & 1711 \\ 
			n = 750 & $n^{1/4}$ & 82 & 87 & 98 & 131 & 173 & 518 \\ 
			& $n^{1/3}$ & 99 & 101 & 116 & 154 & 195 & 564 \\ 
			& $n^{1/2}$ & 177 & 181 & 196 & 234 & 288 & 704 \\ 
			& $n^{2/3}$ & 448 & 446 & 472 & 520 & 585 & 1064 \\ 
			& $n^{3/4}$ & 809 & 789 & 845 & 894 & 955 & 1495 \\ 
			& $n$ & 3364 & 3222 & 3453 & 3350 & 3557 & 4236 \\ 
			n = 1000 & $n^{1/4}$ & 98 & 99 & 118 & 161 & 216 & 709 \\ 
			& $n^{1/3}$ & 122 & 130 & 147 & 198 & 255 & 809 \\ 
			& $n^{1/2}$ & 266 & 260 & 282 & 340 & 407 & 1121 \\ 
			& $n^{2/3}$ & 750 & 781 & 810 & 863 & 970 & 1765 \\ 
			& $n^{3/4}$ & 1502 & 1469 & 1524 & 1585 & 1708 & 2559 \\ 
			& $n$ & 7613 & 7557 & 6842 & 6756 & 7018 & 7687 \\ 
			\bottomrule
		\end{tabular}
		\caption{Average computation time (in seconds) for global survival stacking with various grid choices.}
		\label{tab:rate runtimes}
	\end{table}

	\subsection{Performance under prospective sampling with non-proportional hazards (Scenarios 1 and 2)}
	
\begin{figure}
		\centering
		\includegraphics[width=\linewidth]{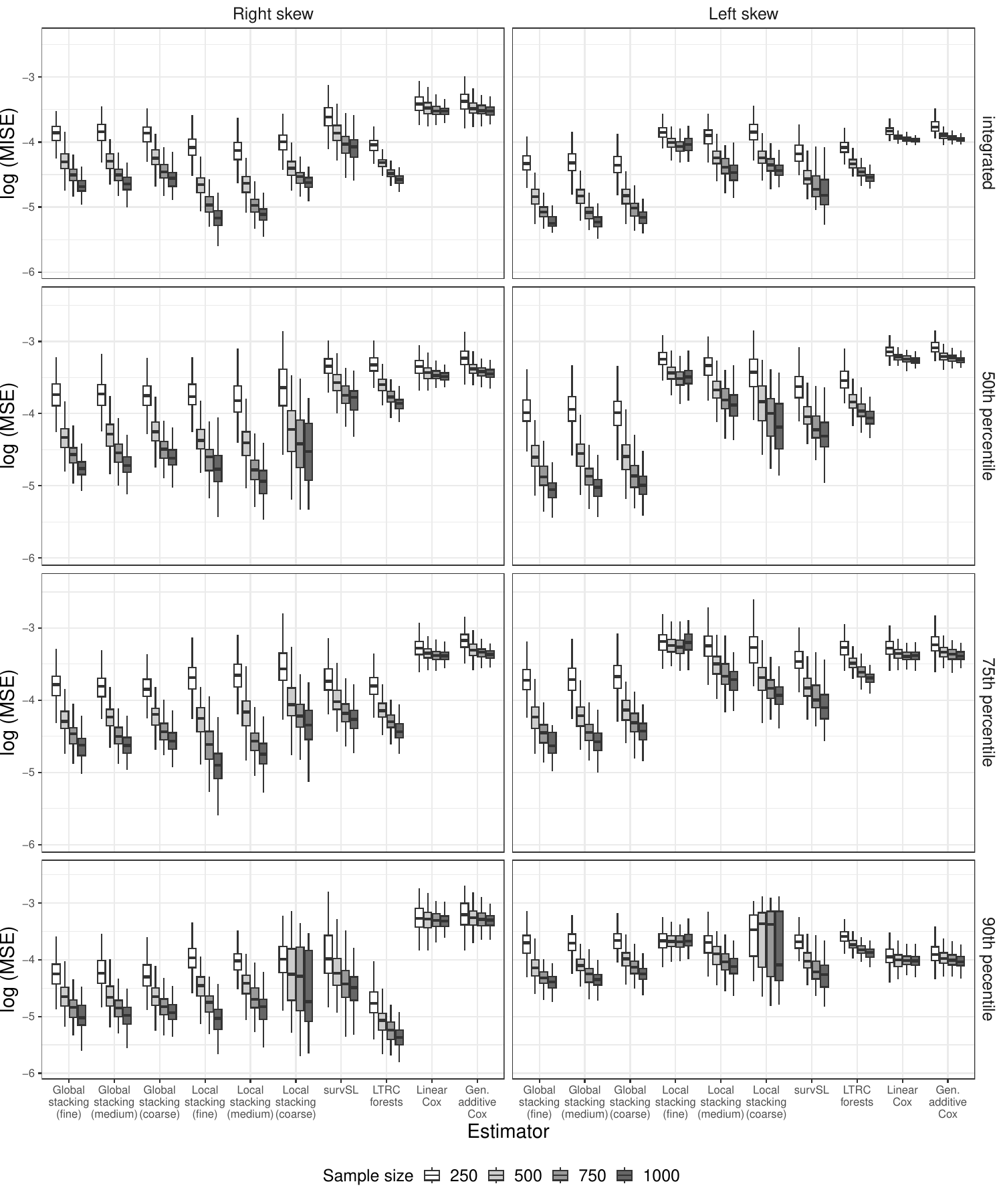}
		\vspace{-0.5cm}
		\caption{Performance of conditional survival estimators with right-censored data (Scenario 1). The methods compared were global survival stacking, local survival stacking, survival Super Learner, random forests, a main-terms linear Cox proportional hazards model with Breslow baseline hazard estimator, and a main-terms generalized additive Cox proportional hazards model with Breslow baseline hazard estimator. Time grids are based on quantiles of observed follow-up times (global stacking) or observed event times (local stacking). The fine grid corresponds to every observed time, the medium grid to 40 cutpoints, and the coarse grid to 10 cutpoints. From top to bottom, rows correspond to MISE and to MSE at 50th, 75th, and 90th percentiles of observed event times. Each boxplot represents 100 simulation replicates.}
		\label{fig:RC integrated supp}
	\end{figure}

\begin{figure}
		\centering
		\includegraphics[width=1\linewidth]{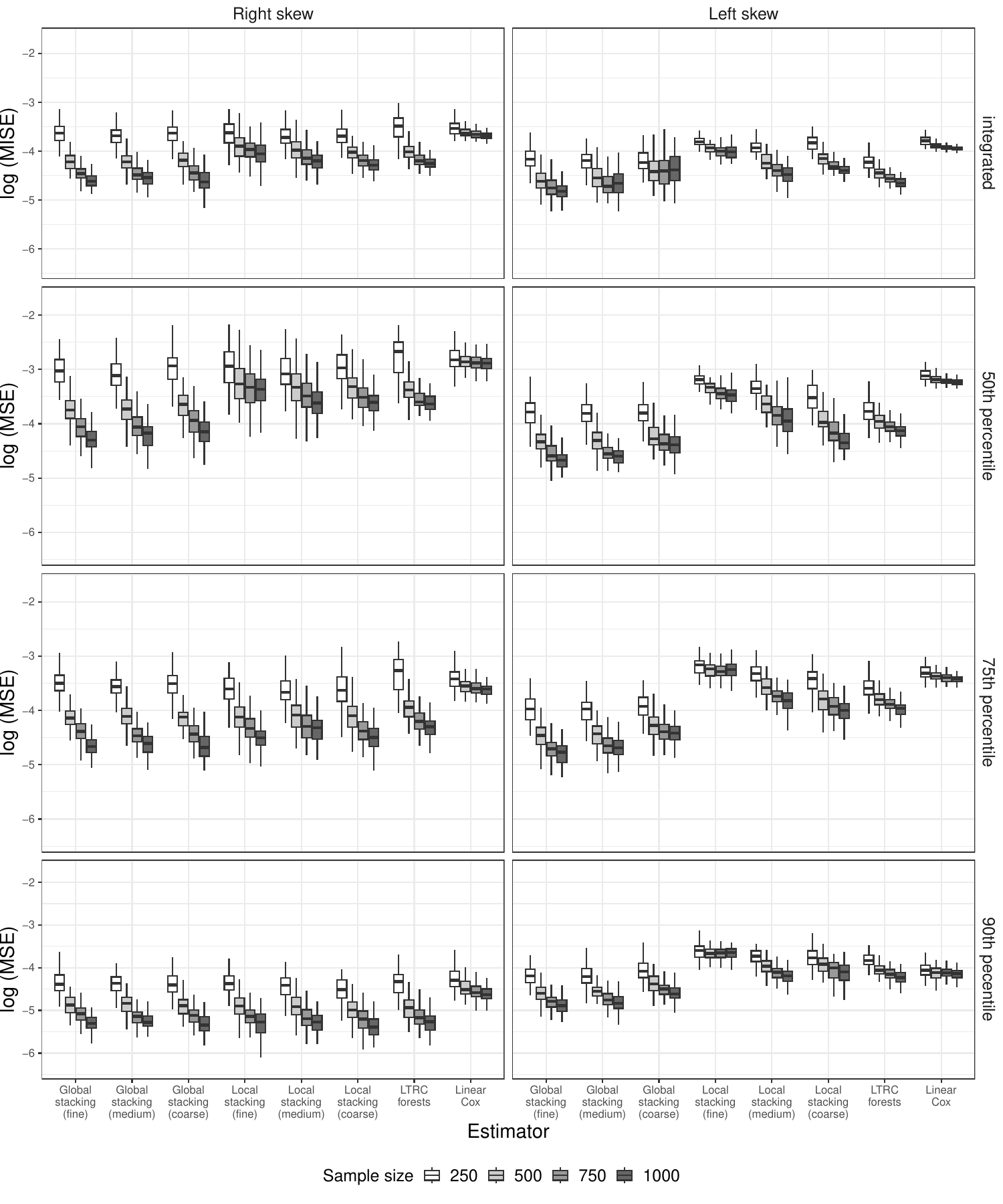}
		\vspace{-0.5cm}
		\caption{Performance of conditional survival estimators with left-truncated, right-censored data (Scenario 2). The methods compared were global survival stacking, local survival stacking, random forests, and a main-terms linear Cox proportional hazards model with Breslow baseline hazard estimator. Time grids are based on quantiles of observed follow-up times (global stacking) or observed event times (local stacking). The fine grid corresponds to every observed time, the medium grid to 40 cutpoints, and the coarse grid to 10 cutpoints. From top to bottom, rows correspond to MISE and to MSE at 50th, 75th, and 90th percentiles of observed event times. Each boxplot represents 100 simulation replicates.}
		\label{fig:LTRC integrated supp}
	\end{figure}

	These simulations were performed under Scenarios 1 and 2, as described in Section \ref{subsec:simulation study} of the main text. Figures \ref{fig:RC integrated supp} and \ref{fig:LTRC integrated supp} display the full results, including mean integrated squared error (MISE) over the interval $[0,100]$ and MSE at landmark times corresponding to the 50\textsuperscript{th}, 75\textsuperscript{th}, and 90\textsuperscript{th} percentiles of observed event times. Global survival stacking performs well overall and is generally not sensitive to the fineness of regression time grids $\mathcal{C}_1$ and $\mathcal{C}_0$. 
	
	\subsection{Performance under retrospective sampling (Scenario 3)}
	
\begin{figure}
		\begin{center}
			\includegraphics[width=1\linewidth]{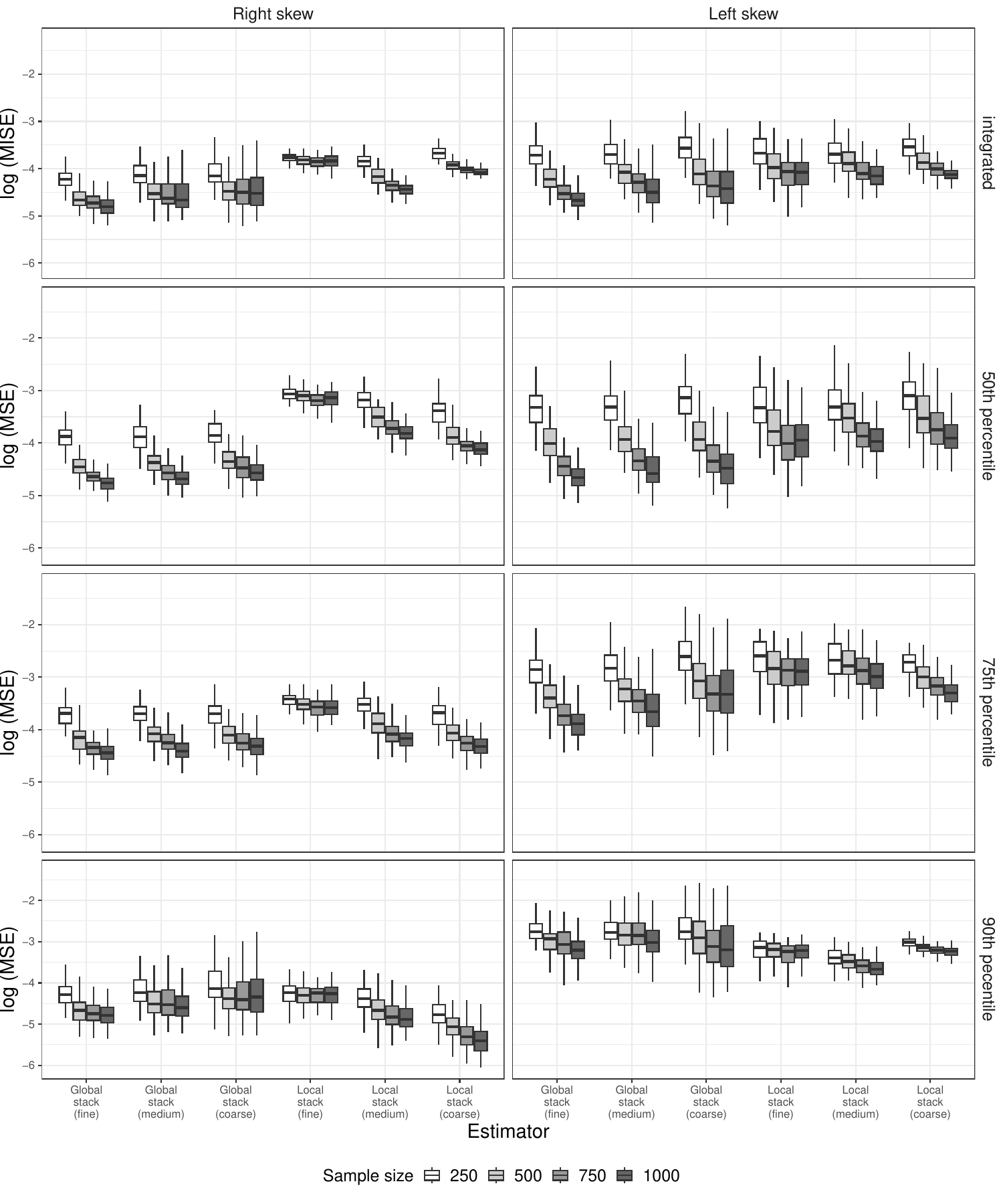}
		\end{center}
		\vspace{-0.5cm}
		\caption{Performance of conditional survival estimators with right-truncated data (Scenario 3). The methods compared were global survival stacking and local survival stacking. Time grids are based on quantiles of observed follow-up times (global stacking) or observed event times (local stacking). The fine grid corresponds to every observed time, the medium grid to 40 cutpoints, and the coarse grid to 10 cutpoints. From top to bottom, rows correspond to MISE and to MSE at 50th, 75th, and 90th percentiles of observed event times. Each boxplot consists of 100 simulation replicates.}
		\label{fig:RT integrated}
	\end{figure}

	For the retrospective simulation study, data were generated as described in Section \ref{subsec:simulation study} of the main text. Only observations with $Y \leq W$ were sampled. There was no censoring in the retrospective study design. Figure \ref{fig:RT integrated} shows the results of the retrospective simulation study. The results are similar as those of the prospective study with left truncation, with global survival stacking demonstrating consistent overall performance. 
	
	\subsection{Performance under proportional hazards (Scenario 4)}
	
	We evaluated the performance of global and local survival stacking when the data satisfied the proportional hazards assumption. The covariate vector $X$, censoring variable $C$, and study entry variable $W$ were generated in the same manner as in the primary numerical experiments described in Section \ref{sec:simulations} of the main text. Next, we set the baseline cumulative hazard function to be that of a random variable $100Z_3$, where in the right-skewed setting $Z_3$ was a Beta(2,3) random variable and in the left-skewed setting $Z_3$ was a Beta(3,2) random variable. Given covariate vector $X = x$, we multiplied this baseline cumulative hazard function by $c(x) = \exp\left\{\tfrac{1}{2}\left(x_1 + x_2 + x_3 + x_4 + x_5\right)\right\}$. For each simulated observation, this yielded a conditional cumulative hazard function satisfying the proportional hazards assumption. Next, we generated $n$ samples from an Exponential(1) distribution, yielding random variables $\xi_1, \xi_2, \ldots, \xi_n$. We applied the inverse of the conditional cumulative hazard functions above to each of the $\xi_i$, resulting in simulated observations distributed according to the specified conditional cumulative hazard functions. We considered the prospective setting with left truncation and 25\% censoring rate. We evaluated performance in the same manner as described in Section \ref{sec:simulations} of the main text. We compared global survival stacking, local survival stacking, and the Cox model with Breslow baseline hazard estimator. 
	
\begin{figure}
		\centering
		\includegraphics[width=\linewidth]{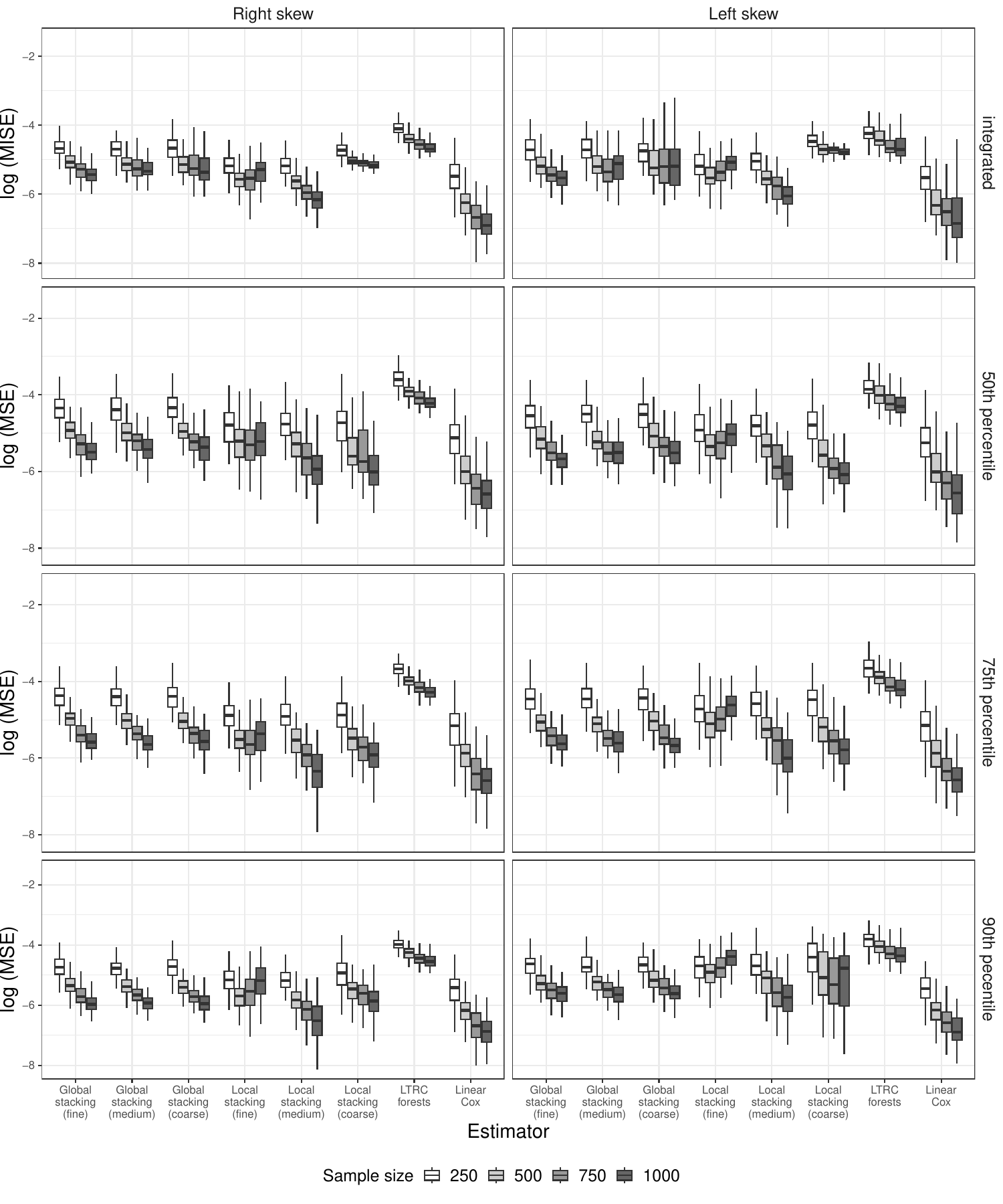}
		\vspace{-0.5cm}
		\caption{Performance of conditional survival estimators with right-censored, left-truncated data generated under a proportional hazards model (Scenario 4). The methods compared were global survival stacking, local survival stacking, random forests, and the main-terms Cox proportional hazards model with Breslow baseline hazard estimator. Time grids are based on quantiles of observed follow-up times (global stacking) or observed event times (local stacking). The fine grid corresponds to every observed time, the medium grid to 40 cutpoints, and the coarse grid to 10 cutpoints. From top to bottom, rows correspond to MISE and to MSE at 50th, 75th, and 90th percentiles of observed event times. Each boxplot consists of 100 simulation replicates.} 
		\label{fig:ph ltrunc}
	\end{figure}

	We display the results for the proportional hazards simulation in Figure \ref{fig:ph ltrunc}. The Cox model, which in this case was correctly specified, yields the best overall performance across all metrics. Among the machine learning approaches, local stacking on a 40 cutpoint grid performs the best by a modest margin, and global survival stacking demonstrates good performance as well.  As in the primary empirical results in Section \ref{sec:simulations} of the main text, local stacking is more sensitive to grid size choice. Local stacking on a grid of all observed event times tends to show increasing estimation error beyond a sample size of 500. 
	
	\subsection{Performance when events are observed at discrete times (Scenario 5)}
	
	For the discrete-time numerical experiments, we generated $X$, $T$, $C$, and $W$ in the same manner as in the primary numerical experiments described in Section \ref{sec:simulations} of the main text. For $m$ the desired number of times in the discrete-time grid, we divided the interval $[0,100]$ into $m$ equally sized intervals $I_1, I_2, \dots, I_m$. For all $Y$ falling in $I_j$, we set $\tilde{Y}$ equal to the right endpoint of interval $I_j$ and used $\tilde{Y}$ as the observed follow-up time. In this way, while the distribution of $T$ was continuous, $\tilde{Y}$ was observed on a discrete time scale. Likewise, for all $W$ falling in $I_j$, we set $\tilde{W}$ equal to the left endpoint of interval $I_j$. We considered the prospective setting with left truncation and 25\% censoring rate. We evaluated performance in the same manner as described in Section \ref{sec:simulations} of the main text and compared the performance of global and local stacking on grids of all observed follow-up and event times, respectively. We used the product integral form for global stacking. We included the main-terms Cox model as a comparator. 
	
\begin{figure}
		\centering
		\includegraphics[width=1\linewidth]{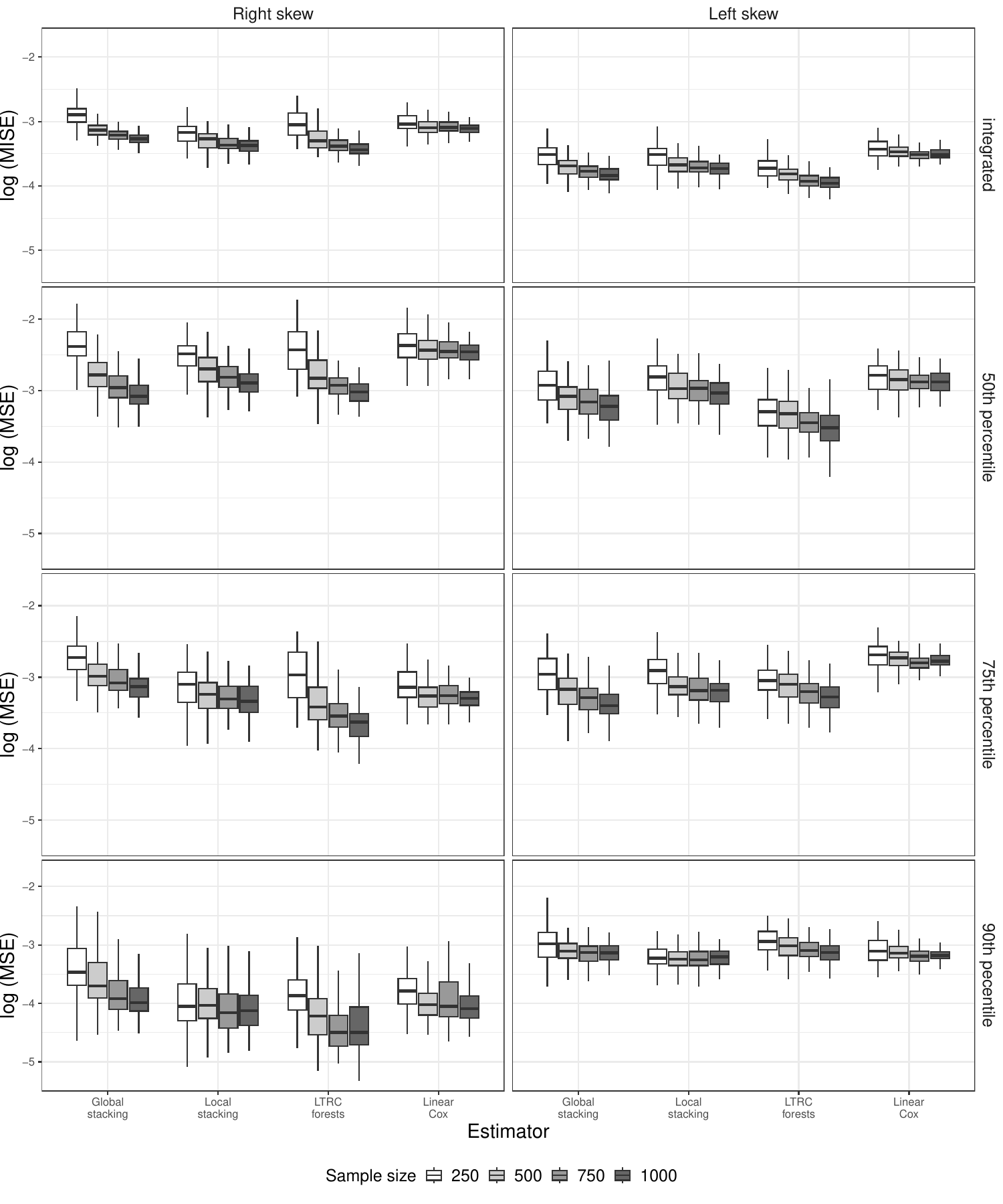}
		\vspace{-0.5cm}
		\caption{Performance of conditional survival estimators with right-censored, left-truncated data observed on a discrete grid of 10 time-points (Scenario 5). The methods compared were global survival stacking, local survival stacking, random forests, and the main-terms Cox proportional hazards model with Breslow baseline hazard estimator. Global and local survival stacking were implemented using a grid of every observed follow-up time (global) or every observed event time (local). From top to bottom, rows correspond to MISE and to MSE at 50th, 75th, and 90th percentiles of observed event times. Each boxplot consists of 100 simulation replicates.} 
		\label{fig:discrete ltrunc 10}
	\end{figure}

\begin{figure}
		\centering
		\includegraphics[width=1\linewidth]{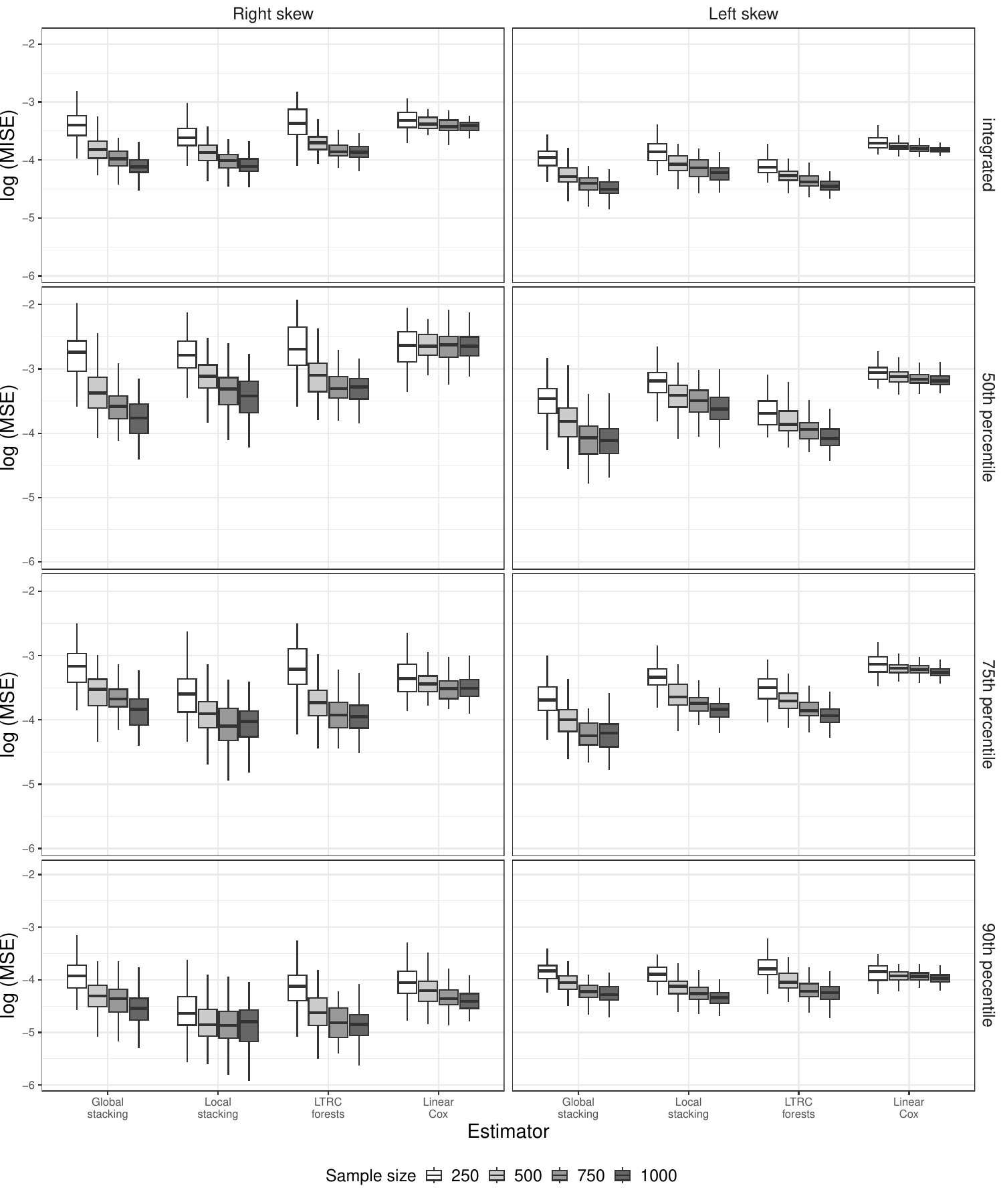}
		\vspace{-0.5cm}
		\caption{Performance of conditional survival estimators with right-censored, left-truncated data observed on a discrete grid of 20 time-points (Scenario 5). The methods compared were global survival stacking, local survival stacking, random forests, and the main-terms Cox proportional hazards model with Breslow baseline hazard estimator. Global and local survival stacking were implemented using a grid of every observed follow-up time (global) or every observed event time (local). From top to bottom, rows correspond to MISE and to MSE at 50th, 75th, and 90th percentiles of observed event times. Each boxplot consists of 100 simulation replicates.} 
		\label{fig:discrete ltrunc 20}
	\end{figure}

\begin{figure}
		\centering
		\includegraphics[width=1\linewidth]{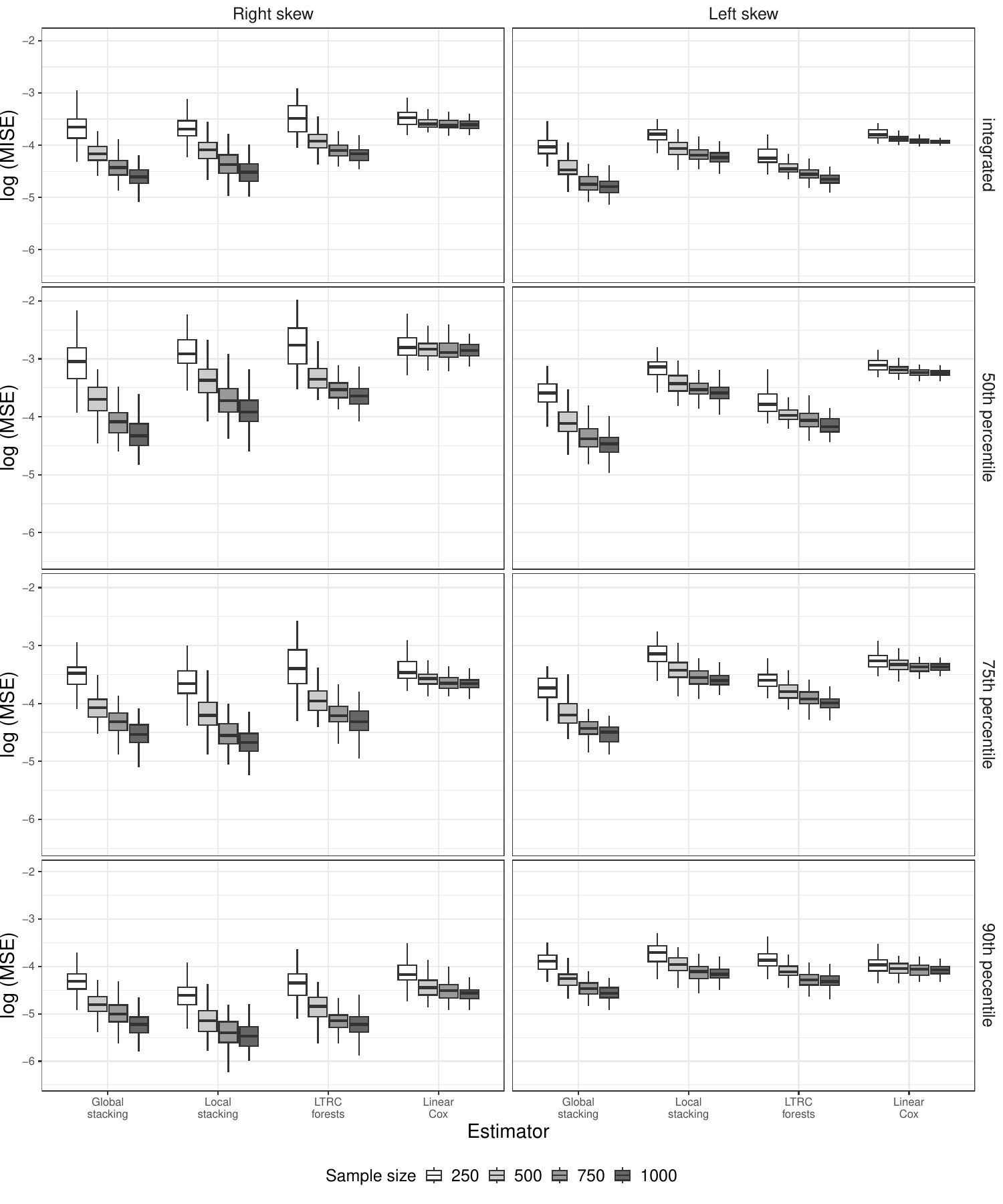}
		\vspace{-0.5cm}
		\caption{Performance of conditional survival estimators with right-censored, left-truncated data observed on a discrete grid of 50 time-points (Scenario 5). The methods compared were global survival stacking, local survival stacking, random forests, and the main-terms Cox proportional hazards model with Breslow baseline hazard estimator. Global and local survival stacking were implemented using a grid of every observed follow-up time (global) or every observed event time (local). From top to bottom, rows correspond to MISE and to MSE at 50th, 75th, and 90th percentiles of observed event times. Each boxplot consists of 100 simulation replicates.} 
		\label{fig:discrete ltrunc 50}
	\end{figure}

	We display the results for the discrete-time experiment with 10 intervals in Figure \ref{fig:discrete ltrunc 10}, with 20 intervals in Figure \ref{fig:discrete ltrunc 20}, and with 50 intervals in Figure \ref{fig:discrete ltrunc 50}. With 10 and 20 intervals, the overall performance of global and local stacking are similar. With 50 intervals, global survival stacking generally outperforms local survival stacking in the left-skewed setting, and the two perform similarly in the right-skewed setting. The MSE and MISE for global stacking are similar in the 50 interval setting as in the continuous-time setting. 
	
	\subsection{Computational considerations}
	
	In order to benchmark the computational burden of different estimators, we simulated samples of size 500 in the prospective study design without left truncation under the left-skewed data-generating mechanism. We fit each estimator as described above and generated conditional survival function estimates for a test data set of size 100 on an evenly spaced grid of times from $t = 0.1$ to $t = 100$. The computational benchmarking simulations were run on an Amazon Web Services EC2 \texttt{r6a.large} instance with 2 vCPUs and 16GB memory. There were 100 simulation replicates for each estimator. 
	
	Table \ref{tab:computation time} displays the results of this experiment. Global survival stacking was slower than alternative methods, and its speed is highly dependent on the size of the grid used in the pooled binary regression.
	
	\begin{table}
		\centering
		\begin{tabular}{l c c} \toprule
			Estimator & Mean runtime (s) & Std. dev. runtime (s)\\ \midrule
			Global stacking (all times grid)&998&36.0\\
			Global stacking (40 cutpoint grid)&222 &43.2\\
			Global stacking (10 cutpoint grid)&129&1.1\\
			Local stacking (all times grid)&493&21.7\\
			Local stacking (40 cutpoint grid)&52&0.8\\
			Local stacking (10 cutpoint grid)&20&0.3\\
			survSuperLearner&61&1.8\\
			LTRC forests &60&1.9\\
			Linear Cox&0.03&0.002\\
			Gen. additive Cox &4.6&0.14\\\bottomrule
		\end{tabular}
		\caption{Computation time for conditional survival estimators from numerical experiments.}
		\label{tab:computation time}
	\end{table} 
	
	\subsection{Comparison of survival function mappings in global survival stacking}
	
	When the product integral is discretized, the differential of the cumulative hazard is a probability and must lie in $[0,1]$. Our method may yield an estimated differential that lies outside of $[0,1]$, leading to survival function estimates that are negative, particularly in the tails of the distribution of $Y$. The exponential form protects against this potential issue and is analogous to exponentiating the negative Nelson-Aalen cumulative hazard estimate \citep{Fleming1984}. We note that in settings without truncation, $S_{n,p}$ naturally respects the $[0,1]$ bounds. When the distribution function of $T$ is continuous, we expect minimal differences in performance between the two forms of the survival function estimator. However, because the exponential mapping from hazard to survival function only holds mathematically if $T$ has a continuous distribution, it is not clear if it should perform as well as the product integral form when the hazard is evaluated on a grid of times. 
	
	We performed a simulation study to compare the two forms (product integral and exponential) of our estimator in the prospective setting with left truncation and right censoring. Both estimators used a grid of 40 cutpoints evenly spaced on the quantile scale. Data were generated as in the other prospective settings, and performance was again evaluated using MISE and MSE at three landmark times. In addition to assessing performance, we also recorded the proportion of estimated survival probabilities in the test data that fell outside the interval $[0,1]$.
	
\begin{figure}
		\centering
		\includegraphics[width=1\linewidth]{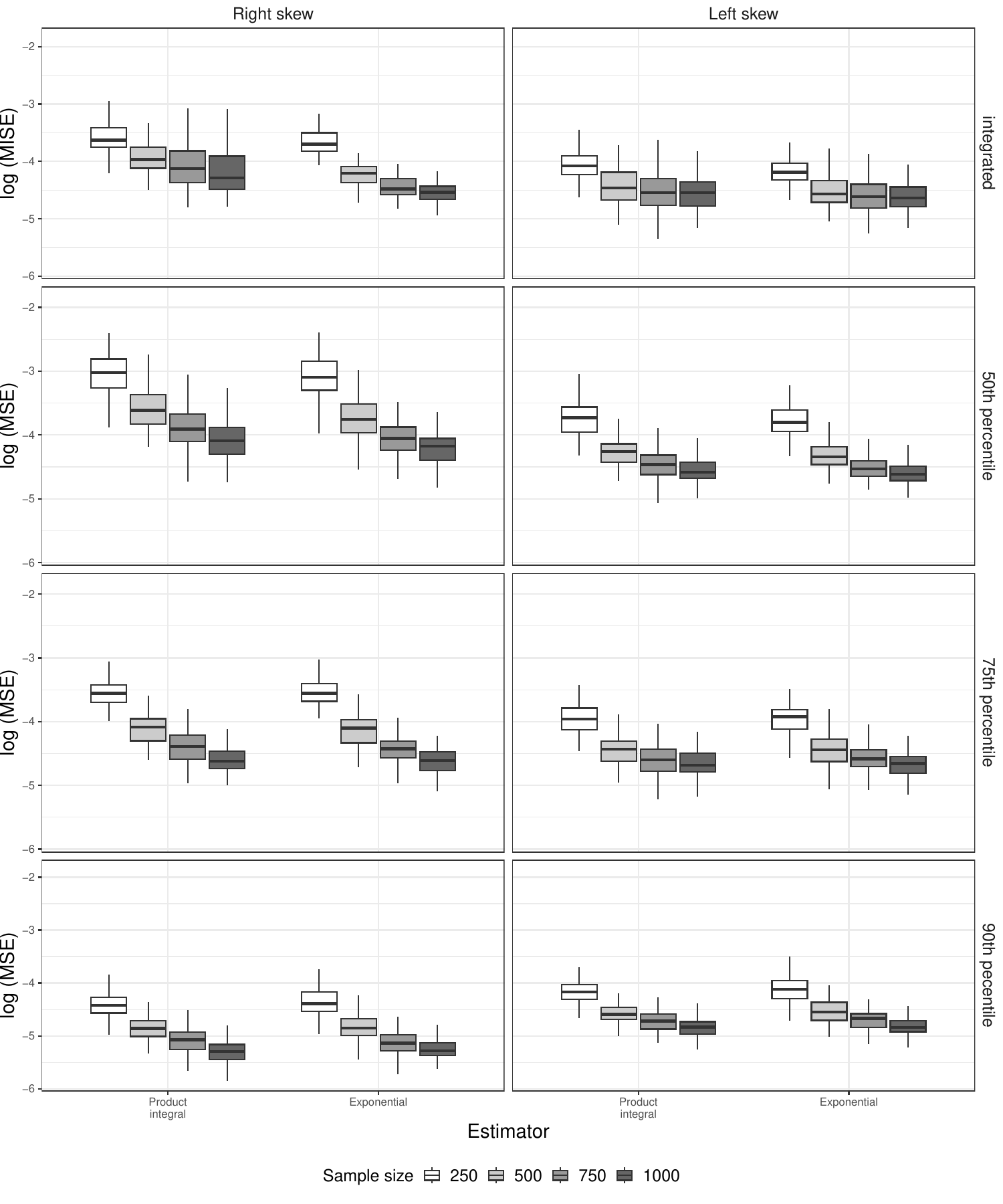}
		\vspace{-0.5cm}
		\caption{Performance of different forms of the global survival stacking estimator in the prospective study design with left truncation and right censoring. The two forms are based on the mappings from hazard to survival function (product integral and exponential). Each boxplot represents 100 simulation replicates.} 
		\label{fig:formcompare}
	\end{figure}

	The overall performance of global survival stacking appears insensitive to the choice of survival function mapping (Figure \ref{fig:formcompare}). For the product integral form, between 0.6\% and 1.5\% of estimated survival probabilities fell outside the unit interval, depending on the training data sample size (Table \ref{tab:incompat}). For the exponential form, none of the survival function estimates fell outside the unit interval. When the distribution of $T$ is continuous, we recommend using the exponential form to protect again potential issues arising in a particular sample.
	
	\begin{table}
		\centering
		\begin{tabular}{l c c} \toprule 
			Estimator & Training sample size & Percent of estimates outside $[0,1]$\\ \midrule 
			Exponential & 250 & 0 \\
			& 500 & 0\\
			& 750 & 0\\
			& 1000 & 0\\
			Product integral & 250 & 0.6\%\\
			& 500& 1.3\%\\
			& 750 & 1.5\%\\
			& 1000 & 1.4\%\\\bottomrule
		\end{tabular}
		\caption{Percentage of estimated survival probabilities falling outside $[0,1]$ using two forms of the global survival stacking estimator in the prospective study design with left truncation and right censoring.}
		\label{tab:incompat}
	\end{table} 
	
	\subsection{Comparison of grid choices for $G$}\label{subsec:W grid}
	
	In the implementation of global survival stacking in Section \ref{sec:simulations}, we based the grids $\mathcal{C}_1$ and $\mathcal{C}_0$ used for pooled binary regression on the distribution of observed event times and observed censoring times, respectively. These grids were used to estimate $(F_1, G_1)$ and $(F_0, G_0)$. Empirically, we found that using quantiles of the distribution of $Y$ in order to estimate $G_1$ and $G_0$ resulted in improved performance compared to using quantiles of the distribution of $W$. We describe those empirical results here. For clarity, we use $\mathcal{C}_{Y, \delta}$ to denote the grid used for $F_\delta$ and $\mathcal{C}_{W, \delta}$ to denote the grid used for $G_\delta$.
	
	In this experiment, we generated data as in Scenario 2. We considered six versions of global survival stacking, based on whether the grids for $F$ and $G$ were shared (i.e., $\mathcal{C}_{Y,\delta} = \mathcal{C}_{W,\delta}$) or different (i.e., $\mathcal{C}_{Y,\delta}$ based on the distribution of observed follow-up times, and $\mathcal{C}_{W,\delta}$ based on the distribution of observed entry times) and on grid fineness (all observed times, 40 cutpoints evenly spaced on the quantile scale, or 10 cutpoints evenly spaced on the quantile scale). We evaluated performance as described in the main text.
	
\begin{figure}
		\centering
		\includegraphics[width=1\linewidth]{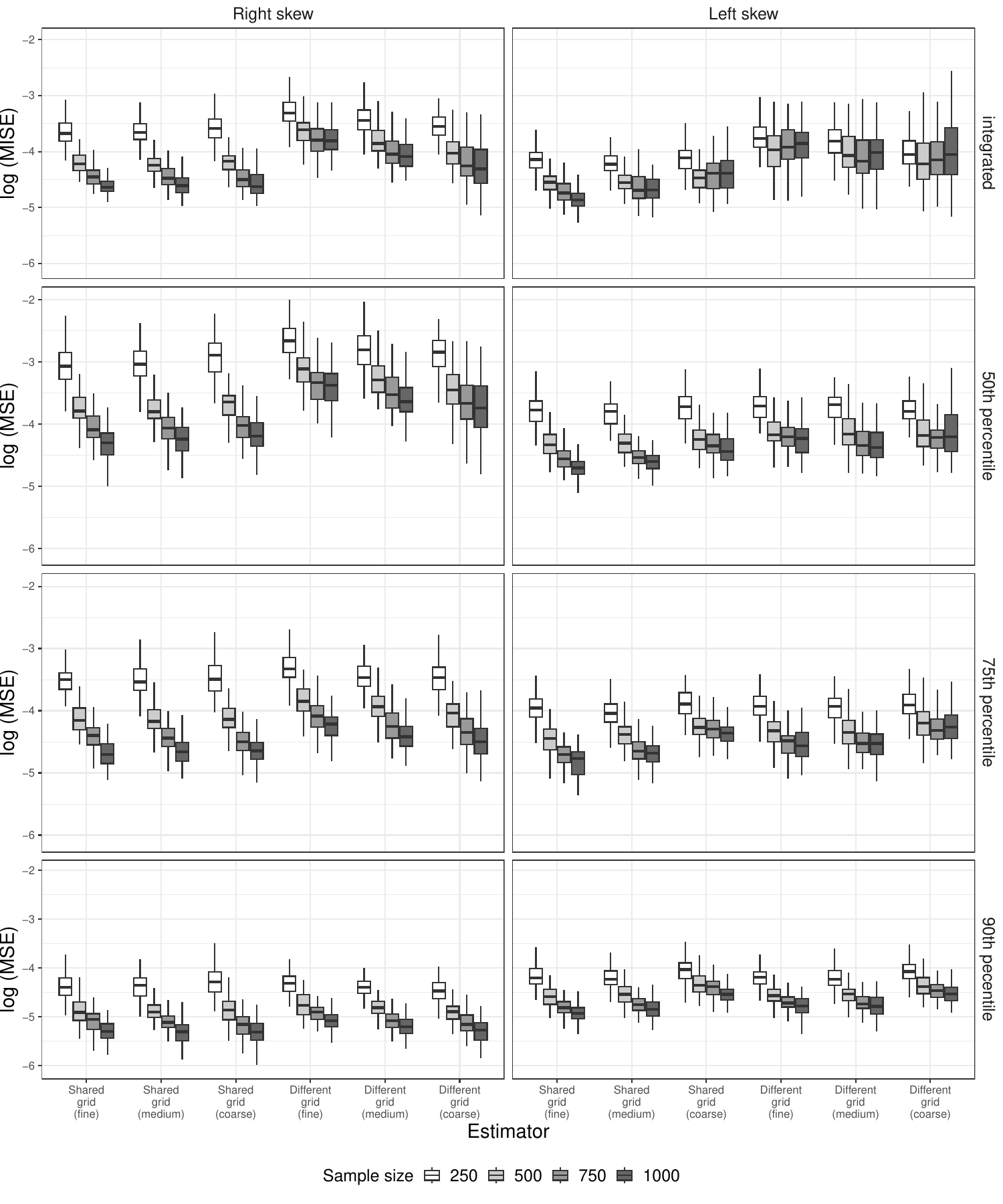}
		\vspace{-0.5cm}
		\caption{Performance of the global survival stacking estimator with various choices of time grid for pooled binary regression in the prospective study design with left truncation and right censoring. Estimators differ by both grid fineness and by whether or not the grid used to estimate $G_1$ and $G_0$ is the same as that used to estimate $F_1$ and $F_0$. The fine grid corresponds to every observed time, the medium grid to 40 cutpoints, and the coarse grid to 10 cutpoints. Each boxplot represents 100 simulation replicates.}
		\label{fig:W grid compare}
	\end{figure}

	Figure \ref{fig:W grid compare} shows the results of this experiment. For any given grid size choice, the shared grid approach demonstrates equal or better performance compared to the different grid approach. The phenomenon is particularly pronounced for the fine time grid.
	
	\section{Details on publicly available datasets}\label{sec:data details}
	
	We describe the publicly available survival datasets analyzed in Section \ref{sec:real data performance} of the main text. 
	
	\noindent{\textbf{FLCHAIN:}} The Assay of Serum-Free Light Chain study investigated the relationship between serum-free light chain and mortality in residents of Olmstead County \citep{Kyle2006}. We used eight features for prediction: age, sex, calendar year of sample collection, serum-free light chain kappa portion, serum-free light chain lambda portion, free light chain group, serum creatinine, and an indicator of monoclonal gammopathy diagnosis. After removal of individuals with missing data, the dataset consisted of 6542 individuals. This dataset is available in the \texttt{survival} package \citep{survival-package}. 
	
	\noindent{\textbf{GBSG:}} The German Breast Cancer Study Group data is derived from a 1984-1989 trial of patients with node-positive breast cancer \citep{Schumacher1994}. The outcome of interest was recurrence-free survival time, with seven features of interest: hormone therapy, age, menopausal status, tumor size, tumor grade, number of positive nodes, progesterone receptor positivity, and estrogen receptor positivity. We used dummy variables for tumor grade, which consists of three categories. After removal of individuals with missing data, the dataset consisted of 684 individuals. It is available in the \texttt{survival} package \citep{survival-package}. 
	
	\noindent{\textbf{METABRIC:}} This dataset was produced by the Molecular Taxonomy of Breast Cancer International Consortium \citep{Curtis2012}. The outcome of interest was mortality, and the features of interest included expression of four different genes (MKI67, EGFR, PGR, and ERBB2), as well as five clinical features (hormone treatment, radiotherapy, chemotherapy, estrogen receptor positivity, and age at diagnosis). This dataset consisted of 1904 individuals, after removal of individuals with missing data. It is available in the \texttt{DeepSurv} software package \citep{Katzman2018}. 
	
	\noindent{\textbf{NWTCO:}} The National Wilms' Tumor Study investigated the relationship between time to tumor relapse and several prognostic variables, including two types of histology \citep{Dangio1976}. We included five features: local histology, central histology, age, disease stage, and an indicator of whether the individual was a participant in NWTS 3 or 4. We used dummy variables for disease stage, which consists of four categories. This dataset consisted of 4028 individuals and is available in the \texttt{survival} package \citep{survival-package}.  
	
	\noindent{\textbf{SUPPORT:}} The Study to Understand Prognoses and Preferences for Outcomes and Risks of Treatments investigated the relationship between clinical outcomes among seriously ill hospitalized adults \citep{Knaus1995}. For our analysis, the outcome of interest was mortality, with 14 features of interest: sex, age, race, number of comorbidities, blood pressure, heart rate, respiration, white blood cell count, temperature, serum creatinine, serum sodium, dementia diagnosis, diabetes diagnosis, and cancer diagnosis. Dummy variables were used for race (five categories) and cancer (three categories). After removal of individuals with missing data, the dataset consisted of 8873 individuals. This dataset is available on the Vanderbilt Biostatistics website. 
	
\end{document}